\documentstyle[12pt]{article}
\begin{document}
\begin{center}
{\Large
       QUANTUM CHROMODYNAMICS AND MULTIPLICITY DISTRIBUTIONS}

{\large                    I.M. Dremin}

P.N.Lebedev Physical Institute, Moscow 117924, Russia

                    Contents
\end{center}

  1. Introduction        \\
  2. Definitions and notations   \\
  3. Phenomenology                 \\
     3.1. KNO--scaling and $F$--scaling  \\
     3.2. Conventional distributions       \\
          3.2.1. Poisson distribution        \\
          3.2.2. Negative binomial distribution   \\
          3.2.3. Fixed multiplicity                 \\
     3.3. Some models                                 \\
  4. Equations of quantum chromodynamics                \\
  5. Gluodynamics                                         \\
     5.1. Approximate solutions of equations with fixed coupling
          constant and the shape of the KNO--function       \\
     5.2. Higher order approximations with running coupling constant \\
  6. Approximate solutions of QCD equations with running coupling constant  \\
  7. Exact solutions of QCD equations with fixed coupling constant
\\
     7.1. First moments and the ratio of average multiplicities in gluon
          to quark jets
\\
     7.2. Higher order moments and widths of distributions in gluon and
          quark jets
 \\
  8. Experiment
\\
  9. Evolution of distributions with decreasing phase space -  \\
     intermittency and fractality     \\
 10. Brief discussion of other QCD effects                                  \\
 11. Conclusions
\\
%\newpage
\begin{abstract}
Quantum chromodymamics (QCD) approach to the problem of multiplicity
distributions in high energy particle collisions is described. The solutions
of QCD equations for generating functions of the multiplicity distributions
in gluon and quark jets are presented for both fixed and running coupling
constants. The newly found characteristics very sensitive to distribution
shapes is discussed. The predictions are confronted to experimental data.
Evolution of the multiplicity distributions with decreasing phase space windows
is considered and discussed in connection to the notions of intermittency and
fractality. Some other QCD effects are briefly described.
\end{abstract}

\Section{Introduction}

\noindent Quantum chromodynamics (QCD) has been treated for a long time as
the theory of strong interactions. Its numerous successes in describing the
static properties of hadrons (especially, those of heavy quarkonia),
the symmetry features of their interactions,
the sum rules are very impressive. Discovery of the asymptotic freedom of
quantum chromodynamics has led to theoretical foundation of the formerly
phenomenological parton model and has opened the way for usage of the
perturbation theory when applied to hadron processes with high transferred
momenta where quarks and gluons play the role of partons [1-5]. Certainly,
the transition of quarks and gluons to experimentally accessible hadrons at
 the final stage of evolution should be considered, and we are still unable
to treat it in a unique way since the problem of confinement has not been
solved
in the framework of quantum chromodynamics even though lattice calculations
tell us
that it is an inherent property of QCD. The simplified estimates show, however,
that either this stage does not change drastically the final results or its
impact does not depend strongly on energy and therefore can be estimated from
other processes at different energies. Phenomenologically, the distributions of
partons and hadrons seem remarkably similar somehow. In such a situation, the
study of the partonic stage
 of the cascade becomes of uppermost importance because the final properties
of multiple production of hadrons at high energies are determined to a great
extent by the partonic cascade.

The distribution of inelastic events according to the number of produced
particles (for the sake of brevity it is called the multiplicity distribution)
is one of the most important features revealing the dynamics of the
interaction.
Phenomenological approaches to its description originate usually from the
simplified ideas about particle emission by several sources and exploit some
distributions widely used in probability theory (see, e.g., [6]). Among them,
the negative binomial distribution is one of the most popular distributions
because it describes reasonably well the experimental data for various
reactions
in wide energy intervals, when its parameters are fitted, though there is some
discrepancy at the highest accessible energies. The attractive feature
of the negative binomial distribution is the KNO--scaling at asymptotically
high energies, i.e. at average multiplicity tending to infinity. According to
the KNO--hypothesis (called by the first letters of the names of its
authors [7]), the multiplicity distributions depend on the ratio of the number
of particles to the average multiplicity only (it is explained in more detail
below). In general features, that property has been confirmed by experiment too
(probably, except some data at highest energies).

What does quantum chromodynamics tell us about multiplicity distributions?
Some attempts to find an answer to it have been undertaken in those papers
which
constitute the main content of the present review paper. It appears that
quantum
chromodynamics when treated in higher order approximations at the partonic
level has been able to predict some very delicate features of multiplicity
distributions which happen to be valid qualitatively for hadrons as well.
Before delving into the details of the conclusions, let us point out some
"underwater stones" and describe in brief the history of the problem.

First of all, one should always keep in mind that quantum chromodynamics
provides conclusions about parton (quark and gluon) distributions but not about
final hadrons as it has been mentioned already. One has to use additional
assumptions to get from them the knowledge of experimentally accessible values.
One of them is the hypothesis about the local parton-hadron duality [8]
which claims that one should just renormalize the parton distributions
without changing their shape to get hadron distributions. It originates
from ideas of "soft" preconfinement [9] when partons group in colourless
clusters without disturbing the initial spectra. Phenomenological models of
hadronization have been attempted in Monte Carlo versions of inelastic
processes
and, in most cases, they support the approximate property of the local
parton-hadron duality even though there exist some quantitative differencies
from that hypothesis as it is described below.

Another problem, tightly connected to the first one, is the limitations of
the perturbation theory analysis in a definite approximation. Formally
speaking, one can apply the perturbation theory just when the coupling constant
is very small. That condition is fulfilled in quantum chromodynamics for
extremely large transferred momenta only. In each process, however, the energy
of cascading partons degrades during their evolution and one has to take into
account
in a proper way soft partons, their recoil due to interaction and
energy-momentum
conservation laws. All these factors used to be neglected in the lowest order
approximation when the processes with high gradient of energies and of emission
angles at each stage of the evolution are considered only (the so-called
double logarithmic approximation). An account of soft partons and of strict
angular ordering is done in the next terms of the perturbation theory series
such as modified leading logarithm approximation and next-to-next-to leading
terms. The recoil effects and conservation laws are taken into account at
that stage too.

In most cases those corrections are well under control and constitute tens
per cents of the main term. In spite of rather low integral contribution,
they are, nevertheless, very important on the qualitative level drastically
changing the whole situation in the region where the corresponding functions
are small. For example, they become crucial for proper restoration of the
multiple production processes. It reveals itself mathematically as a new
expansion parameter equal to the product of the coupling constant (to
be more precise, of its square root) and the rank of the moment of the
distribution. Thus it is large at large ranks, i.e. at high multiplicity.
We shall describe these problems in more detail in Sections 3--5.

That is why the very first results on multiplicity distributions of partons
in quantum chromodynamics were obtained using double logarithmic approximation
(see reviews in [5, 10]). They are both impressive and discouraging. First,
they
proclaim the asymptotical KNO--scaling of the distributions which does not
depend on the value of the coupling constant at all [11]. One is tempted to
speculate about somewhat more general meaning of the result. However, it fails
to be valid in higher-order approximations. The energy increase
of the average multiplicity depends on the coupling constant. It is faster than
any
logarithmic function and slower than any power-like one (if the running
coupling constant is used) what agrees quite well with experimental findings.
At the same time, the shape of the KNO--function contradicts to any
experimental
distributions because it is much wider than any of them. Just recently, it
became
possible to resolve the problem [12, 13] by proper account of the higher
order effects mentioned above. In any case, one can state now that the
agreement with experiment is achieved at the qualitative level, at least.
Moreover, some qualitative predictions of the perturbative quantum
chromodynamics are unexpectedly well suited for "soft" hadronic processes as
well. From one side, it puzzles even though one recognizes that higher
order corrections should take into account ever softer partons in a consistent
way. From another side, it implies, probably, the more general nature of
soft and hard processes than it has been imposed on various theoretical
schemes and persuades to reconsider our approaches to the origin of effects
under consideration (leading, for example, to the experimentally observed
multiplicity distributions). Besides, the newly found characteristics
prompted by solutions of QCD equations are extremely sensitive to tiny
details of the multiplicity distributions.
The modern state of affairs for the multiplicity distributions in quark and
gluon jets described by quantum chromodynamics is the main concern of the
present article (see Sections 5--8).

What concerns the average multiplicities, many discussions are devoted to
the value of the ratio of average multiplicities in gluon and quark jets. Its
initial value obtained in the double logarithmic approximation is equal to 9/4.
It strongly exceeds all experimental estimates. The simplest corrections
reduce that value by about ten per cents. Even stronger decrease of it has
been predicted by the exact solution of the equations for generating functions
in the case of fixed coupling and by the higher-order approximations with
running coupling constant. Its derivation and correspondence to experimental
data is discussed in Sections 6, 7. The energy dependence of the average
multiplicity is considered there also.

For the sake of completeness, one should mention other interesting facts
of inelastic interactions of high energy particles which are successfully
described (and sometimes predicted) by quantum chromodynamics. Effects
predicted by quantum chromodynamics are very exquisite, sometimes unexpected
and always extremely instructive. When studying the multiplicity distributions,
one is eager to ask about their behaviour not only in the total phase
space but in its smaller subregions too. It is well known that these studies
are very popular nowadays. They are related to intermittency phenomenon and to
fractality of particle distributions within the small phase space volume. It
results from enlarged fluctuations in such phase space regions (for the latest
review see [14]). Those features are caused by the relative widening of the
multiplicity distributions in smaller phase space volumes. It gives rise to the
increase of their moments in a power-like manner directly revealing properties
of intermittency and fractality. Such tendencies have been experimentally
observed. Quantum chromodynamics describes the increase of the moments,
relates the intermittency exponents (or fractal dimensions) directly to
the anomalous dimension and clearly depicts the region of applicability of
those regularities indicating the scales at which one should take into
account the running coupling constant or consider it as a fixed one. We
describe these results briefly in Section 8.

Quantum-mechanical origin of the interacting partons reveals itself in various
interference effects. They lead to the hump-backed plateau of rapidity
distributions, to the correlations of partons in energies and azimuthal
angles, to the string (or drag-)-effect in the three-jet events and in
production of
heavy bosons and lepton pairs at large transverse momenta, to the suppression
of forward production of accompanying particles in processes with heavy
quarks. We describe them
briefly in Section 9. Unfortunately, all peculiarities of interactions with
nuclei as well as interactions of
polarized quarks are not even mentioned in the paper to keep it in a reasonable
shape. They deserve a separate publication.

My main concern here are the multiplicity distributions and related characteris
tics. I apologize to all authors of numerous papers on multiplicity distributio
ns whose contribution has not been mentioned. My only excuse stems from the
intention to describe just quantum chromodynamics approach to the subject. Even
there, some omissions could happen unintentionally, unfortunately.

\Section{Definitions and notations}

\noindent The distribution of the number of particles produced in high energy
inelastic events is called the multiplicity distribution and is given by
the formula
\begin{equation}
P_{n} = \frac {\sigma_{n}}{\sum_{n=0}^{\infty}\sigma_{n}} ,    \label{1}
\end{equation}
\noindent where $\sigma_{n}$ is the cross section of $n$-particle production
processes (the so called topological cross section) and sum is over all
possible values of $n$ so that
\begin{equation}
\sum_{n=0}^{\infty}P_{n} = 1 .          \label{2}
\end{equation}
Sometimes it is more convenient to replace the multiplicity distribution
by its moments, i.e. by another set of numbers obtained from it by the definite
algorithm. All such sets can be obtained from the so called generating function
defined by the formula
\begin{equation}
G(z) = \sum_{n=0}^{\infty }P_{n}(1+z)^{n}  ,                     \label{3}
\end{equation}
which substitutes the analytical function in place of the set of numbers
$P_{n}$.

In what follows we shall often use the (normalized) factorial moments $F_{q}$
and
cumulants $K_{q}$ determined from the generating function $G(z)$ by the
relations
\begin{equation}
F_{q} = \frac {\sum_{n} P_{n}n(n-1)...(n-q+1)}{(\sum_{n} P_{n}n)^{q}} =
\frac {1}{\langle n \rangle ^{q}} \frac {d^{q}G(z)}{dz^{q}}\vert _{z=0} ,
\label{4}
\end{equation}
\begin{equation}
K_{q} = \frac {1}{\langle n \rangle ^{q}} \frac {d^{q}\ln G(z)}{dz^{q}}
\vert _{z=0}, \label{5}
\end{equation}
where
\begin{equation}
\langle n \rangle = \sum_{n=0}^{\infty }P_{n}n              \label{6}
\end{equation}
is the average multiplicity. The expression for $G(z)$ can be rewritten as
\begin{equation}
G(z) = \sum _{q=0}^{\infty } \frac {z^q}{q!} \langle n \rangle ^{q} F_{q}
\;\;\;\; ( F_0 = F_1 = 1 ) ,   \label{7}
\end{equation}
\begin{equation}
\ln G(z) = \sum _{q=1}^{\infty } \frac {z^q}{q!} \langle n \rangle ^{q} K_{q}
\;\;\;\; ( K_1 = 1 ).    \label{8}
\end{equation}
The distribution $P_{n}$ and its ordinary moments $C_{q}$ are derived from the
generating function $G(z)$ according to the formulae
\begin{equation}
P_{n} = \frac {1}{n!} \frac {d^{n}G(z)}{dz^{n}}\vert _{z=-1} ,
\label{9}
\end{equation}
\begin{equation}
C_{q} = \frac {\sum _{n=0}^{\infty }P_{n}n^{q}}{\langle n \rangle ^{q}} =
\frac {1}{\langle n \rangle ^{q}} \frac {d^{q}G(\e^{z}-1)}{dz^{q}}\vert _{z=0}
. \label{10}
\end{equation}
All the moments are connected by definite relations which can be easily derived
from their definitions by the generating function. For example, the factorial
moments and cumulants are related to each other by the identities
\begin{equation}
F_{q} = \sum _{m=0}^{q-1} C_{q-1}^{m} K_{q-m} F_{m} ,              \label{11}
\end{equation}
which are nothing else as the relations between the derivatives of a function
and of its logarithm at the point where the function itself equals 1. Here
\begin{equation}
C_{q-1}^{m} = \frac {(q-1)!}{m!(q-m-1)!} = \frac {\Gamma (q)}{\Gamma (m+1)
\Gamma (q-m)} = \frac {1}{mB(q,m)}     \label{12}
\end{equation}
  are the binomial coefficients, and $\Gamma $, B denote the gamma-  and
 beta-functions, correspondingly. Thus there are just numerical
  coefficients in recurrent relations (\ref{11}) only and the iterative
solution (well-suited for computer calculation) reproduces all cumulants if
the factorial moments have been given, and vice versa. In that sense, cumulants
and factorial moments are equally suitable. The physical meaning of both of
them
is clearly seen from their definitions if they are presented in the form of
integrals of correlation functions. However we do not write them down
here (for review, see \cite{14}) but refer to the analogous relations in
quantum field theory where formulae similar to (\ref{4}) and (\ref{5}) define
the whole set of Feynman graphs (both connected and disconnected ones) and
the subset of connected diagrams,
correspondingly (see, e.g., \cite{1}). Therefrom it is easy to recognize
that the factorial moments are the integral characteristics of any correlations
among the particles while the cumulants of $q$-th rank correspond to "genuine"
$q$-particle correlations not reducible to the product of lower order
correlations {\footnote This interpretation is valid, however, only for the
moments with the rank smaller than the average multiplicity at given energy
(for more details, see the review paper \cite{14}).}. To be more precise, all
$q$ particles are connected to each other in the $q$-th cumulant and can not be
split into the disconnected groups. One can say that they form the $q$-particle
cluster which is not divided in smaller clusters, in analogy with Mayer cluster
decomposition in the statistical mechanics.

It is a common feature of distributions in particle physics that their
factorial moments and cumulants increase fastly at large ranks $q$.
That is why it is convenient to consider their ratio
\begin{equation}
H_{q} = \frac {K_q}{F_q}       ,      \label{13}
\end{equation}
which behaves in a more "quiet" way at high ranks $q$ emphasizing at the
same time all typical qualitative peculiarities of cumulants as functions of
their rank $q$.

Using the definition of factorial moments (\ref{4}) one can easily derive
their relation to the ordinary moments $C_q$ of the same and lower ranks.
The coefficients depend on the mean multiplicity so that, for example,
\begin{equation}
F_{2} = \frac {\langle n(n-1)\rangle }{\langle n \rangle ^2} = C_2 - \langle
n \rangle ^{-1} .     \label{14}
\end{equation}
It complicates the whole matter since one should recalculate them at any
given energy if the $F$--scaling persists. That is why we do not use the
ordinary
moments in what follows.
In asymptotics, the ordinary and factorial moments coincide, however.

One should keep in mind that the generating function contains the same
physical information  as the multiplicity distribution. It is true also for
unnormalized moments and for their ratio. For normalized moments, one should
define the average multiplicity at a given energy. Let us point out that the
higher rank moments lay emphasis on higher multiplicity events. The
multiplicities
$n\geq q$ contribute to the factorial moment of the (integer) rank $q$ as is
seen from (\ref{4}). If the distribution is cut off at some $n=n_{max}$, all
factorial moments with the rank $q>n_{max}$ are equal to zero while they are
positive at smaller $q$. The cumulants may be either positive or negative.

Up to that moment, without mentioning it, we have assumed that the rank of the
moment is an integer positive number. However, the definitions (\ref{4}),
(\ref{5}), (\ref{10}) can be generalized to include non-integer moments
\cite{15}.
It is easily done by rewriting the factorial moments as
\begin{equation}
F_{q} = \frac {1}{\langle n\rangle ^{q}}\sum _{n=0}^{\infty }P_n \frac
{\Gamma (n+1)}{\Gamma (n-q+1)} .       \label{15}
\end{equation}
It is valid at any real value of the rank $q$.
The same formula is obtained by applying differentiation of real order $q$
(i.e. by using the fractional calculus) to the generating function.

According to the generalized differentiation rule one gets a fractional
derivative of any (real) order if the whole set of ordinary derivatives
of integer order is known \cite{16}:
\begin{equation}
D_{z}^{q}G(z) = \sum_{m=0}^{\infty }\frac {(1+z)^{m-q}G^{(m)}(-1)}{\Gamma
(m-q+1)} ,   \label{16}
\end{equation}
where $G^{(m)}(-1)$ are the derivatives of the integer (positive) order $m$ of
the function $G(z)$, defined by eq. (3), at $z=-1$. The generalized definition
of factorial moments for any real (both positive and negative) non-integer and
integer $q$ can be written as:
\begin{equation}
F_{q} = \frac {1}{\langle n \rangle ^q} D_{z}^{q}G(z)\vert _{z=0} =
\frac {1}{\langle n \rangle ^q}\sum _{m=0}^{\infty }\frac {G^{(m)}(-1)}
{\Gamma (m-q+1)} .     \label{17}
\end{equation}
The formulae (\ref{9}), (\ref{15}), (\ref{17}) correspond to each other, i.e.
the experimental definition of factorial moments by the formula (\ref{15})
is equivalent to its theoretical definition by (\ref{17}) as the fractional
derivatives of the generating function. At integer ranks one gets the
previously
used formulae. That is why the generalized (to non-integer ranks) moments are
known as fractional moments. Their use could help distinguish various
distributions as we discuss later.

The situation with cumulants is more complicated. It is straightforward to
define [17] them theoretically as
\begin{equation}
K_{q} = \frac {1}{\langle n \rangle ^q} D_{z}^{q}\ln G(z)\vert _{z=0} = \frac
{1}{\langle n \rangle ^q}\sum _{m=0}^{\infty }\frac {(\ln G(z))^{(m)}\vert
_{z=-1}}{\Gamma (m-q+1)} . \label{18}
\end{equation}
However, the relation between factorial moments and cumulants becomes much
more complicated than formula (\ref{11}) and impractical to use. We should
keep in mind that our aim is not to calculate the cumulants themselves but
to find the characteristics of multiplicity distributions which are most
sensitive to their shape. Therefore, it has been proposed to use the so called
"analytically continued" cumulants denoted by $K_{q}^{(a)}$ and defined by
the recursion relations at any real rank $q$
\begin{equation}
F_{q} = \sum _{m=0}^{[q-1]}(mB(q,m))^{-1}K_{q-m}^{(a)}F_{m} . \label{19}
\end{equation}
The sum is up to the integer part of $q-1$. As is easily seen from formulae
(\ref{11}), (\ref{12}), it can be used for any value of $q$, integers
including.
It is convenient both for experimentalists and theorists after some additional
convention. The relation to the
"true" cumulants defined by (\ref{18}) has been lost, however.

\Section{Phenomenology}

\Subsection{KNO--scaling and $F$--scaling}

\noindent One of the most successful assumptions about the shape of the
multiplicity distributions at high energies is the hypothesis that the energy
dependence is determined completely by the behaviour of the average
multiplicity in such a way that the distribution $P_n$ may be represented as:
\begin{equation}
P_{n} = \frac {1}{\langle n \rangle}f(\frac {n}{\langle n \rangle}) .
\label{20}
\end{equation}
That property has been called the KNO--scaling according to the names of its
authors \cite{7} who proposed it relying on Feynman plateau of rapidity
distributions. The normalization condition (\ref{2}) leads to
\begin{equation}
\int _{0}^{\infty }f(x)dx = 1 .   \label{21}
\end{equation}
It is clear that the ordinary moments of the KNO--distribution (\ref{20})
do not depend on energy and are just the functions of their rank $q$
\begin{equation}
C_{q} = \int _{0}^{\infty }x^{q}f(x)dx = \const (E) .   \label{22}
\end{equation}
The factorial moments of the distribution are energy dependent tending to
constant values at asymptotically high energies since they differ from the
ordinary moments by lower-order correlation terms suppressed by inverse of
the average multiplicity to the corresponding power. Constancy of the
factorial moments will be called $F$--scaling. It coincides with KNO--scaling
in asymptotics. As is clear from definitions (\ref{3}), (\ref{7}), (\ref{8}),
the generating function depends on the average multiplicity in both cases only.

In quantum chromodynamics with fixed coupling constant (see section 7),
$F$--scaling is preferred. However, the difference
from KNO--scaling is usually neglected since the theoretical calculations
are often performed for asymptotically high energies. The preasymptotic
correction terms in the second moment have been considered in \cite{18}.

In the double logarithmic
approximation the equations for factorial moments are independent of energy and
of coupling constant at all. The corresponding function $f(x)$ decreases
exponentially \cite{5} at large $x$ :
\begin{equation}
f(x) \sim 2C(Cx-1+\frac {1}{3Cx}+...)\e ^{-Cx} ; \;\;\;\;\; Cx\gg 1  ,
\label{23}
\end{equation}
where $C\approx 2.553$. At low $x$ it behaves as:
\begin{equation}
f(x) \sim x^{-1}\exp (-\ln ^{2}x/2) .    \label{24}
\end{equation}
Though the very appearance of the KNO--scaling and its independence on
coupling constant in the lowest approximation are by themselves the great
success of the perturbative quantum chromodynamics \cite{11}, the shape of the
scaling function (\ref{23}) does not fit experimentally obtained shapes.
Experiment favours the shapes which are much narrower than it is prescribed by
(\ref{23}), (\ref{24}). The corrections of the modified leading logarithmic
approximation indicate that the resulting form should become less wide
\cite{19}.
It happens that the higher order terms reduce the width of $f(x)$ but it
depends on coupling constant now. Those problems are treated in Sections 5--7.

Up to now, we implicitly assume that we are treating the multiplicity
distributions in the total phase space. It is reasonable to ask the question
about their evolution if some restrictions are imposed on the region of the
phase space under investigation. In particular, one can study such
distributions
in ever smaller rapidity intervals contained within the total interval. In that
case, the moments of the distribution become, in general, the functions of the
size of the interval beside their ranks (for review, see \cite{14}). Their
behaviour is often related to the notions of intermittency and fractality
discussed briefly in Section 9.

\Subsection{Conventional distributions}

We shall consider three distributions where analytical expressions for
generating
functions and all moments can be derived \cite{20},\cite{21}. They will serve
us the "starting points" for further discussion of QCD-distributions. First, we
shall describe the moments of the integer rank, and then show what happens with
arbitrary (fractional, negative, complex rank) moments.

\Subsubsection{Poisson distribution}

The presence of correlations in a process is conventionally described by the
difference between its typical distribution and the Poisson distribution
which is written as
\begin{equation}
P_{n} = \frac {\langle n \rangle ^n}{n!} \e ^{-\langle n \rangle } . \label{25}
\end{equation}
The generating function is  (see (\ref{3}))
\begin{equation}
G(z) = \exp (\langle n \rangle z)           \label{26}
\end{equation}
and acccording to (\ref{4}), (\ref{5}) one gets
\begin{equation}
F_{q} = 1 , \;\;\; K_{q} = H_{q} = \delta _{q1} .   \label{27}
\end{equation}
Therefore the measure of correlations could be defined as the difference
between
$F_q$ and 1 or between $K_q$  ($H_q$) and 0. There is exact $F$--scaling and
asymptotical KNO--scaling.

The fractional (in general, complex rank) factorial moments of the Poisson
distribution [15] are
\begin{equation}
F_{q} = \frac {\e ^{-\langle n\rangle }}{\langle n \rangle ^{q}\Gamma (1-q)}
\Phi (1, 1-q; \langle n \rangle ) ,  \label{28}
\end{equation}
where $\Phi $ is the degenerate confluent hypergeometric function. At integer
positive values of $q$ they are equal to 1, as it should be, and oscillate with
an
amplitude depending on $q$ and $\langle n \rangle $ in the intervals between
the integer ranks as it is shown in Fig. 1 ( the curve $e$).

The cumulants [17] are
\begin{equation}
K_{q} = \frac {q}{\langle n \rangle ^{q-1}\Gamma (2-q)}  \label{29}
\end{equation}
and ratios are
\begin{equation}
H_{q} = \frac {q}{1-q} \frac {\langle n \rangle \e ^{\langle n \rangle }}
{\Phi (1, 1-q; \langle n \rangle )} .    \label{30}
\end{equation}
One gets (\ref{27}) from (\ref{28}), (\ref{29}), (\ref{30}) at integer positive
$q$. The amplitude of oscillations of moments decreases fastly with the
increase of the average multiplicity. Actually, they originate from quite
simple properties of the gamma-function in the denominator.

At large average multiplicity all factorial moments tend to 1 in the whole
complex plane $q$. However, cumulants tend to zero only at Re$q>1$
 and increase in absolute value at
 Re$q<1$ with the increase of mean multiplicity what can be used for analysis
 of distributions in the total phase space at high energies.  Qualitative
features  of that kind are typical for other distributions considered below.

\Subsubsection{The negative binomial distribution (NBD)}

The negative binomial distribution deserves special attention because it has
been actively used during several last years to fit the experimental
multiplicity
distributions and has been rather successful. In particular, there exists the
widely spread opinion that it describes almost all inelastic processes at high
energies (except of the data at highest available energies, i.e. for
$e^{+}e^{-}$ at
91 GeV --DELPHI \cite{22}, OPAL \cite{23}-- and for proton-antiproton
interactions at energies from 200 to 900 GeV --UA5 \cite{24}). For NBD we have
\begin{equation}
P_{n} = \frac {\Gamma (n+k)}{\Gamma (n+1)\Gamma (k)} \left ( \frac {\langle n
\rangle }{k}\right )^{n} \left (1+\frac {\langle n \rangle }{k}\right )^{-n-k}
,
\label{31}
\end{equation}
where $k$ is an adjustable parameter with a physical meaning of the number of
the
independent sources. The Bose-Einstein distribution is a special
case of NBD with $k=1$. The Poisson distribution is obtained from (\ref{31}) in
the limit $k\rightarrow \infty $. The generating function is
\begin{equation}
G(z) = \left ( 1-\frac {z\langle n \rangle }{k}\right ) ^{-k} ,  \label{32}
\end{equation}
and the (integer rank) moments are
\begin{equation}
F_{q} = \frac {\Gamma (k+q)}{\Gamma (k) k^q} ,   \label{33}
\end{equation}
\begin{equation}
K_{q} = \frac {\Gamma (q)}{k^{q-1}} ,   \label{34}
\end{equation}
\begin{equation}
H_{q} = \frac {\Gamma (q)\Gamma (k+1)}{\Gamma (k+q)} = kB(q, k) . \label{35}
\end{equation}
 The factorial moments increase faster
than exponent with $q$ at a fixed value of $k$. The cumulants are steeply
decreasing at small $q$ till they reach a minimum at $q\approx k$ and start
increasing at larger $q$. They always stay positive. Let us note that the
product
of several generating functions of negative binomial distributions with
different parameters leads also to positive cumulants since the unnormalized
total
cumulant is just the sum of unnormalized individual cumulants.
The ratio $H_q$ is positive also and decreases monotonically (as $q^{-k}$ at
large $q$).

In Fig. 2 we show their behaviour by plotting $\ln F_q, \ln K_q$ and $\ln H_q$
vs $q$ for $k$ = 5 and 10. Since $P_n$ is narrower at higher $k$, the slower
rise of $F_q$ for $k$ = 10 compared to that for $k$ = 5 is expected. The
dependence
of $K_q$ on $k$ is more pronounced than that of $H_q$. These properties are
more
characteristic of NBD than general and do not reveal themselves in quantum
chromodynamics.

It should be noted that the negative binomial distribution with the fixed
parameter $k$ possesses the $F$--scaling (the moments do not depend on energy)
and the asymptotical KNO--scaling. The KNO-function at large $n$ and fixed
values of $k$ behaves as
\begin{equation]
f(x) = \frac {k^k}{(k-1)!}x^{k-1}\e ^{-kx} .   \label{36}
\end{equation}
The generating function (\ref{32}) is singular at the point $z=k/\langle n
\rangle
\rightarrow 0$ for $\langle n \rangle \rightarrow 0$ and $k=\const $.
Therefore,
we have to deal in the vicinity of the singularity when calculating its
derivatives at $z=0$ (factorial moments). The singularity moves closer to $z=0$
at higher energies.

The general expressions for the moments valid for any
rank $q$ in the whole complex plane are
\begin{equation}
F_{q} = \frac {(kv)^k}{\langle n \rangle ^{q+k}} \frac {F(1, k; 1-q; v)}{\Gamma
(1-q)}  = \frac {F(k, -qw; 1-q; -\langle n \rangle /k)}{\langle n \rangle ^{q}
 \Gamma (1-q)} , \label{37}
\end{equation}
\begin{equation}
K_{q} = \frac {k}{\langle n \rangle ^{q} \Gamma (1-q)} \left [ \frac {v}{1-q}
F(1, 1; 2-q; v) + \ln (kv/\langle n \rangle )\right ] ,  \label{38}
\end{equation}
\begin{equation}
H_{q} = k \left (\frac {\langle n \rangle }{kv}\right )^{k} \frac {\frac
{v}{1-q}
F(1, 1; 2-q; v) + \ln (kv/\langle n \rangle )}{F(1, k; 1-q; v)} ,  \label{39}
\end{equation}
where $v=\langle n \rangle /(\langle n \rangle + k)$.

Oscillations of moments between integer values of $q$ diminish at higher
average
multiplicities (i.e. at higher energies) and at lower values of $k$. It is
shown
in Figs.1(a-d) for $F_q$. They are really very small at high energies. The
oscillations are imposed on fast increase of $F_q$ with $q$.

At negative values of $q$ the moments increase with average multiplicity. In
the
complex plane, their oscillations are noticeable (for example, along lines
parallel to the real axis).

\Subsubsection{Fixed multiplicity distribution}

We consider the case of fixed multiplicity just to show that the behaviour of
moments (even for integer ranks) can drastically differ from the examples
treated above. Besides, it demonstrates how important the role of the selection
procedure in experiment could be. Really, it often happens that the events
with a given multiplicity are chosen for analysis (so called semi-inclusive
events), i.e. one deals with the distribution
\begin{equation}
P_{n} = \delta _{nn_{0}} \;\;\;\;\;\;\; (n_{0} = \const ) .  \label{40}
\end{equation}
Then one gets
\begin{equation}
G(z) = (1+z)^{n_0} .   \label{41}
\end{equation}
Since $\langle n \rangle = n_0 $, we obtain
\begin{equation}
F_{q} = \frac {n_{0}!}{n_{0}^{q} (n_{0}-q)!} = \frac {\Gamma (n_{0})
n_{0}^{1-q}}
{\Gamma (n_{0}-q+1)} ,    \;\;\;\;\;  1<q\leq n_{0} ,   \label{42}
\end{equation}
\begin{equation}
F_{q} = 0 ,     \;\;\;\;\;\;\;\;  q>n_{0} ,    \label{43}
\end{equation}
\begin{equation}
K_{q} = (-n_{0})^{1-q}(q-1)! = (-n_{0})^{1-q}\Gamma (q) ,   \label{44}
\end{equation}
\begin{equation}
H_{q} = (-1)^{1-q}n_{0}B(q, n_{0}-q+1) .   \label{45}
\end{equation}
All factorial moments of the rank higher than $n_{0}$ are identically zero
and one can calculate $H_q$ at $q\leq n_{0}$ only. The typical feature of
that distribution is the alternating signs of integer order cumulants which are
positive at odd values of $q$ and negative at even values. The amplitude of
oscillations decreases when $q$ increases from 1 to $n_{0}$ and then increases
monotonically. The change of the sign (but with a different periodicity) will
be seen in QCD as well. Factorial moments, however, behave differently in the
two cases. They decrease monotonically with $q$ until $n_{0}$ for fixed
multiplicity and increase fastly in QCD.

In Fig.3  we show $F_{q}, K_{q}, H_{q}$ for $n_{0}=10$, and in the insets we
show
$\ln \vert K_q \vert $ and $\ln \vert H_q \vert $ for integer values of $q$.
Straight lines connect just the points at integer values of $q$ and are shown
to guide the eye.

Let us stress that the very existence of the oscillations can be related just
to the selection procedure of the events and, in the case of fixed
multiplicity,
has nothing to do with the dynamics
of the interaction. It is easy to recognize when one chooses, e.g., 10-particle
events only from the set of them with Poisson distribution (or any other). Then
we obtain alternating sign cumulants at integer ranks instead identically equal
to zero. Their amplitude can preserve the memory about the original
distribution
if its normalization has been kept untouched.

The fractional moments calculated according to their definitions at any rank
$q$
(\ref{17}), (\ref{18}) are
\begin{equation}
F_{q} = n_{0}^{1-q}\frac {B(n_{0},1-q)}{\Gamma (1-q)} ,  \label{46}
\end{equation}
\begin{equation}
K_{q} = n_{0}^{1-q}\frac {\psi (1) - \psi (1-q)}{\Gamma (1-q)} ,  \label{47}
\end{equation}
\begin{equation}
H_{q} = \frac {\psi (1) - \psi (1-q)}{B(n_{0},1-q)} .   \label{48}
\end{equation}
For fixed $q$ and $n_{0}\rightarrow \infty $ one gets $F_{q} \rightarrow 1 ,
K_{q}\rightarrow 0, H_{q}\rightarrow 0$.

Let us stress that in all the cases considered the non-integer rank moments
are not obtained by the simple-minded "analytical continuation" according to
the formula (\ref{19}) but are given by the different expressions ( compare
(\ref{27}) with (\ref{28})-(\ref{30}), (\ref{33})-(\ref{35}) with (\ref{37})-
(\ref{39}) or (\ref{42})-(\ref{45}) with (\ref{46})-(\ref{48}) ) which are
derived from the proper definitions (\ref{17}), (\ref{18}).

The common property of oscillations between the integer ranks is the change
from maximum to minimum at each subsequent rank. At the integer points, there
are the knots for Poisson and negative binomial distributions and just maxima
or minima for fixed multiplicity.

Unfortunately, the oscillations between the integer ranks are small at high
multiplicity and can be useful for distributions with low average
multiplicity like those in small phase space volumes considered in Section 9.
At high multiplicity they impose the low-amplitude harmonics on the main
dependence, and may be neglected in the first approximation.

At the same time, the increase of the negative moments with average
multiplicity can be useful for analysis of distributions within the
total phase space.

\Subsection{Some models}

At the very first sight, the theoretical diagram description of multiparticle
production looks completely different in $e^{+}e^{-}$ and hadron-hadron
processes. In the former case, main graphs are of the tree-like type with a
highly-virtual initial parton. In the latter case, one used to consider a
sequence of the multiperipheral type graphs with low virtualities and
rather complicated topology. More general (and unifying) picture emerges from
consideration of strings between the colour
charges in the process of their interaction (Lund model \cite{22,23}, dual
topological model \cite{24,25}, quark-gluon string model \cite{26,27}) and of
final particles clusterization (multiperipheral cluster model \cite{28},
clans \cite{6}, \cite{29} etc.). The multiplicity distributions in the models
are not described usually by a single analytical formula but are formed from
a combination of several distributions. For example, the multiperipheral
model with a single ladder gives rise to the Poisson distribution of the
particle emission centers (resonances, fireballs, clusters, clans ...).
In general, the resulting distribution is obtained by convolution of the
Poisson
distribution of the number of sources with the decay
distribution which can describe experimental data quite well for educated guess
about decay properties. If one chooses the logarithmic distribution of cluster
decay multiplicity then its convolution with Poisson distribution of clusters
produces the negative binomial distribution of final particles multiplicities.
The simultaneous creation of several ladders (or strings) gives rise to a more
complicated shape of the distribution. Sometimes it may be approximated by the
sum of the negative binomial distributions with distinct parameters. As a
result,
the distributions with "shoulders" or "quasi-oscillations" imposed on smooth
curves can be observed. The possible relation of such oscillations with those
of $H_q$ discussed in the present review has been considered in \cite{30}.
 Analogously, a single jet in $e^{+}e^{-}$-annihilation can give rise to the
 negative binomial distribution while the superposition of several jets differs
 from it in a resulting shape \cite{31}. The detailed study of those semi-
 phenomenological models is usually done with Monte Carlo computing.

\Section{Equations of quantum chromodynamics}

The multiparticle production processes are described in quantum chromodynamics
as a result of the interaction of quarks and gluons which leads to creation
of additional quarks and gluons forming the observed hadrons at the very last
stage. The most typical features of the processes are determined by the vector
nature of gluons and by the dimensionless coupling constant. The gluons are
colour charged in distinction to photons which have no electric charge.
Therefore,
they can emit gluons in addition to quark-antiquark pairs. That is why both
quark and gluon jets are considered in quantum chromodynamics as main
objectives.
Their development is described by the evolution equations. The main parameter
of
the evolution is the opening angle of the jet or its transverse momentum. The
subsequent emission of gluons and quarks fills in the internal regions of the
previously developed cones so that they do not overlap (angular ordering). This
remarkable property
can be exploited to formulate the probabilistic scheme for the development of
the jet as a whole. Then its evolution equations remind the well-known
classical
Markovian equations for the "birth--death" (or "mother--daughter") processes.
(The detailed discussion of that approach, based on the coherence phenomenon,
see in \cite{5}).

The system of two equations for the generating functions $G_F$ and $G_G$ of
the quark and gluon jets, correspondingly, are (here A, B, C = F, G) \cite{1,
5}
\begin{equation}
G_{A}(y,z) = \e ^{-w_{A}(y)}z + \frac {1}{2}\sum _{B,C}\int ^{y}dy\prime \int
_{0}
^{1}dx \e ^{-w_{A}(y)+w_{A}(y\prime )}\frac {\alpha_{S}}{2\pi }K  _{A}^{BC}(x)
G_{B}(x,y\prime )G_{C}((1-x),y\prime ) ,  \label{49}
\end{equation}
where $y=\ln p\Theta /Q_0 , p$ is an initial momentum, $\Theta $ is the
opening angle of the jet, $Q_{0}=\const , \alpha _{S}$ is the coupling
constant.
The first term in the right-hand side corresponds to the propagation of the
primary
parton without any evolution that is described by the form-factor $\exp [-w_{A}
(y)]$. The second term shows the creation of two jets $B$ and $C$ with
shares of the primary energy $x$ and $1-x$, correspondingly, after their
production at the vertex $K _{A}^{BC}$ with the evolution parameter $y\prime $
which has been reached by the primary parton without splitting  as is dictated
by the factor $\exp [-w_{A}(y)+w_{A}(y\prime )]$.

Multiplying both sides of the equation by $\exp [w_{A}(y)]$ and differentiating
over $y$, we get rid of all form-factors and obtain the final system of
equations:
\cite{1},\cite{5}
\begin{equation}
G_{G}^{\prime } = \int_{0}^{1}dxK_{G}^{G}(x)\gamma _{0}^{2}[G_{G}(y+\ln x)G_{G}
(y+\ln (1-x)) - G_{G}(y)] + n_{f}\int _{0}^{1}dxK_{G}^{F}(x)\gamma _{0}^{2}
[G_{F}(y+\ln x)G_{F}(y+\ln (1-x)) - G_{G}(y)] ,   \label{50}
\end{equation}
\begin{equation}
G_{F}^{\prime } = \int _{0}^{1}dxK_{F}^{G}(x)\gamma _{0}^{2}[G_{G}(y+\ln x)
G_{F}(y+\ln (1-x)) - G_{F}(y)] ,                                   \label{51}
\end{equation}
where $G^{\prime }(y)=dG/dy , n_f$ is the number of active flavours,
\begin{equation}
\gamma _{0}^{2} =\frac {6\alpha _S}{\pi } ,                \label{52}
\end{equation}
and the kernels of the equations are
\begin{equation}
K_{G}^{G}(x) = \frac {1}{x} - (1-x)[2-x(1-x)] ,    \label{53}
\end{equation}
\begin{equation}
K_{G}^{F}(x) = \frac {1}{4N_c}[x^{2}+(1-x)^{2}] ,  \label{54}
\end{equation}
\begin{equation}
K_{F}^{G}(x) = \frac {C_F}{N_c}[\frac {1}{x}-1+\frac {x}{2}] ,   \label{55}
\end{equation}
where $N_c$=3 is the number of colours, and $C_{F} = N_{c}[1-N_{c}^{-2}]/2
=4/3$ in QCD.

We have omitted the variable $z$ in the generating functions. One should keep
in mind, however, that the derivation of the equations for moments relies
completely on the expansions (\ref{7}), (\ref{8}) when they are inserted into
the above equations and the coefficients in front of terms $z^q$ are compared.

The typical feature of any field theory with the dimensionless coupling
constant
(of quantum chromodynamics, in particular) is the presence of the singular
terms at $x\rightarrow 0$ in the kernels of the equations. They imply the
uneven sharing of energy between newly created jets and play an important role
in jet evolution.

Even though the system of equations (\ref{50}), (\ref{51}) is physically
appealing, it is not absolutely exact, i.e. not derived from first principles
of
quantum chromodynamics. One immediately notices it since, e.g., there is no
four-gluon interaction term which is contained in the lagrangian of QCD.
Such a term would not lead to singular contribution to the kernels and its
omission is justified in lowest orders. Nevertheless, the modified series of
the perturbation theory (with three-parton vertices) is well
reproduced by such equations up to the terms
including two- and three-loop corrections. As shown in \cite{5}, the neglected
terms would contribute at the level of the product of, at least, five
generating
functions. Physical interpretation of the corresponding graphs would lead to
treatment of the "colour polarizability" of jets. There are some problems with
the definition of the evolution parameter, with preasymptotic corrections etc.
(see, e.g., \cite{32}). The above arguments do not
prevent from further detailed studies of higher order corrections to these
equations, and it seems reasonable to learn more about the solutions of the
equations with higher accuracy since there are indications that neglected terms
are not very important.

\Section{Gluodynamics}

It is quite natural to start our studies with the simplest case of
gluodynamics.
There are no quarks in that case, and interactions of gluons are considered
only.
The system of equations (\ref{50}), (\ref{51}) degenerates to the single
equation
\begin{equation}
G^{\prime }(y) = \int _{0}^{1}dxK(x)\gamma _{0}^{2}[G(y+\ln x)G(y+\ln (1-x)) -
G(y)] ,   \label{56}
\end{equation}
where $G(y)\equiv G_{G}(y), K(x)\equiv K_{G}^{G}(x)$. It is the non-linear
integro-
differential equation with shifted arguments in the non-linear term which take
into account the energy conservation. In the lowest order double-logarithmic
approximation, one considers the most singular terms in the kernel $K(x)$ and
inside the square brackets, i.e. $1/x$ in $K$ and $\ln (1-x)\rightarrow 0$,
while $\gamma _{0}^{2}$ is chosen constant {\footnote Somewhat inconsistently,
the running coupling constant is sometimes considered in that approximation
also (see, e.g., \cite{1, 5})}.

\Subsection{Approximate solutions of equations with fixed coupling constant and
the shape of the KNO--function}

Formally speaking, the three assumptions of the double-logarithmic
approximation for the terms under the integral sign in (\ref{56}) are
equivalent
because one neglects non-leading contributions. The detailed analysis of any
of them has been done in many papers \cite{9} - \cite{13},\cite{21},\cite{32} -
 \cite{37} one by one or in some combinations. In most papers the lower
 moments have been treated only, i.e. the average multiplicity and the
 dispersion. It has been noticed that the
 role of conservation laws displayed in shifted arguments of the
generating functions is the most important one. They provide larger
corrections.
It was shown recently \cite{12} that they could be precisely taken into
account.
However, the running property of the coupling constant was disregarded, the
non-singular terms in the kernel were neglected (as well as some other terms)
and
the difference between the coupling constant $gamma _{0}$ (\ref{52}) and
the QCD anomalous dimension $\gamma $ defined as:
\begin{equation}
\langle n \rangle = \exp (\int ^{y}\gamma (y\prime )dy\prime )   \label{57}
\end{equation}
was neglected also. In section 7, we show that equations (\ref{50}), (\ref{51})
possess the exact solutions for fixed coupling constant without any additional
assumptions. Nevertheless, it is instructive to consider this case because
one gets the analytical expression for the KNO--function which clearly reveals
the importance of the conservation laws and differs from the formula (\ref{23})
of the double logarithmic approximation by the smaller width, thus getting much
closer to experimental values.

First, we obtain the system of recurrent equations for factorial moments when
the relations (\ref{7}) are substituted in the equation (\ref{56}) and
the coefficients at $z^q$ are equated in both sides:
\begin{equation}
(q-q^{-1})F_{q} = \gamma \sum _{l=1}^{q-1} C_{q}^{l}B(\gamma l,\gamma (q-l)+1)
F_{q-l}F_{l} .   \label{58}
\end{equation}
That system can be computed {\footnote The exact solution of the system of
equations for quark and gluon jets is given in section 7.2} with initial
conditions $F_{0} = F_{1} = 1$.
In the inset of Fig.4a we show their ratio to the asymptotical solution
\cite{12} of the equations (\ref{58}):
\begin{equation}
F_{q}^{as} = \frac {[\Gamma (1+\gamma )]^{q}}{\Gamma (1+\gamma q)} \frac {2q
\Gamma (q+1)}{C^{q}} .   \label{59}
\end{equation}
One obtains at large $q\gamma $
\begin{equation}
F_{q}\approx \frac {2\mu D^{-q}}{\sqrt {2\pi \gamma }}\Gamma \left ( \frac
{3}{2}
+\frac {q}{\mu }\right ) ,    \label{60}
\end{equation}
where $\mu =(1-\gamma )^{-1}, D=C \gamma ^{\gamma }(1-\gamma )^{1-\gamma }/
\Gamma (1+\gamma )$. The asymptotics of $F_q$ determines the asymptotics of
the KNO--function $f(x)$:
\begin{equation}
f(x)\approx \frac {2\mu ^{2} (Dx)^{3\mu /2}}{x\sqrt {2\pi \gamma }} \exp [-(Dx)
^{\mu }] ,   \;\;\;\;\;\;  (\mu -1)(Dx)^{\mu }\gg 1 .  \label{61}
\end{equation}
It is clear to see that the tail of the distribution at large multiplicities
is suppressed much stronger than in the double logarithmic approximation. One
gets "almost Gaussian" suppression instead the exponent of (\ref{23}) if one
considers the practically important values of $\gamma $ for which $\mu = (1-
\gamma )^{-1}\approx 1.6$. Thus we conclude that conservation laws reduce
drastically the width of the multiplicity distribution. It is demonstrated in
Fig.5 where the modified distribution (which takes into account the behaviour
at
low multiplicities \cite{12}) is compared to the results of the lowest order
QCD and to its fit by the negative binomial distribution with the parameter
$k=7$. Making use of the modified curve in Fig.5, one is able to compute
the "genuine" (with low-multiplicity correction) factorial moments. Their ratio
to the asymptotical solution (\ref{59}) is shown in the main part of Fig.4a.
Comparison of two curves in that Figure reveals quite important influence
of corrections done at low multiplicities. From "genuine" factorial moments,
one can compute cumulants and the "genuine" ratio $H_q$ (the latter one is
shown in Fig.4b and compared to the negative binomial distribution prediction
for $k=7$ and $\langle n \rangle $=30).
One notices the visible difference from the negative binomial distribution in
the ratios $H_q$ while it is hard to see it in distributions shown in Fig.5.
The oscillations of "genuine" $H_q$ contrast its smooth behaviour in negative
binomial distribution. Here they remind somewhat the fixed multiplicity
toy-model considered above. Similar shapes with oscillations of different (!)
periodicity will be discussed in what follows.

\Subsection{Higher order approximations with running coupling constant}

The equation for the generating function in gluodynamics (\ref{56}) can be
solved
in a somewhat different approximation by taking into account all (including
non-singular) terms of the kernel $K$, considering running coupling constant
 $\gamma _{0}$ distinct from the anomalous dimension $\gamma $, and using
the Taylor series expansion for the generating functions in the non-linear term
at large $y$:
\begin{equation}
G(y+\epsilon )\approx G(y) + G^{\prime }(y)\epsilon +\frac {1}{2} G^{\prime
\prime }(y)\epsilon ^{2} + ...     \label{62}
\end{equation}
That approach clearly shows the distinction between various assumptions, their
importance and qualitative effects due to the higher order corrections.

Using (\ref{62}) for the generating functions in the non-linear term of eq.
(\ref{56}), dividing both sides of it by $G(y)$ and differentiating by $y$,
we obtain
\begin{equation}
[\ln G(y)]^ {\prime \prime } = \gamma _{0}^{2}\left [ G(y)-1-2h_{1}G^{\prime }
(y)+\sum _{n=2}^{\infty }(-1)^{n}h_{n}G^{(n)}(y)+\sum _{m,n=1}^{\infty }(-1)^{m
+n}
h_{nm}\left (\frac {G^{(m)}G^{(n)}}{G}\right )^{\prime }\right ] ,  \label{63}
\end{equation}
where
\begin{equation}
h_{1}=11/24, h_{n}=\vert 2-2^{-n}-3^{-n}-\zeta (n)\vert ; \zeta (n)=
\sum _{m=1}^{\infty }m^{-n}, \;\;  n\geq 2;  \label{64}
\end{equation}
\begin{equation}
h_{mn}=\vert \frac {1}{m!n!}\int _{0}^{1}dxK(x)\ln ^{n}x\ln ^{m}(1-x)\vert .
\label{65}
\end{equation}
Leaving two terms in the right hand side, one gets the well known \cite{5}
equation of the double logarithmic approximation which takes into account the
most singular components. The next term with $h_1$ corresponds to the modified
leading logarithm approximation, and term with $h_2$ deals with higher order
corrections. Let us note that we neglect for the moment the dependence of
$\gamma _{0}$ on $y$ in the integral term since it leads to terms of the order
of $\gamma _{0}^{2}$ compared to those written above.

The straightforward solution of the equation (\ref{63}) looks very problematic
even if the terms with $h_1$ and $h_2$ are included in addition to double
logarithmic ones only. However, it is very simple for the moments of
distributions \cite{13} since $G(z)$ and $\ln G(z)$ are the generating
functions for
factorial moments and cumulants, correspondingly. Using the formulae (\ref{7}),
(\ref{8}) in case of $F$--scaling one gets the product $q\gamma $ (and its
derivatives) at each differentiation of those functions because the average
multiplicity is the only $y$-dependent term left. The coefficients in front of
$z^q$ in both sides should be equal. Therefrom one obtains
\begin{equation}
H_{q} = \frac {K_q}{F_q} = \frac {\gamma _{0}^{2}[1-2h_{1}q\gamma +h_{2}(q^{2}
\gamma ^{2}+q\gamma ^{\prime })]}{q^{2}\gamma ^{2}+q\gamma ^{\prime }} .
\label{66}
\end{equation}
The anomalous dimension $\gamma $ is defined by eq. (\ref{57}). The condition
$F_{1}=K_{1}=1$ determines the relation between $\gamma $ and $\gamma _{0}$:
\begin{equation}
\gamma \approx \gamma _{0} - \frac {1}{2}h_{1}\gamma _{0}^{2} + \frac {1}{8}
(4h_{2}-h_{1}^{2})\gamma _{0}^{3} + O(\gamma _{0}^{4}) ,  \label{67}
\end{equation}
which shows that the increase of the average multiplicity with energy is
slower in the modified leading logarithm approximation as compared to the
double
logarithmic approximation because the term with $h_{1}$ is negative (see
(\ref{57})). However,
the higher order terms slightly enlarge it again ($4h_{2}-h_{1}^{2}>0$) but
those corrections are not large. The running property of $\gamma _{0}$ has been
taken into account in (\ref{67}):
\begin{equation}
\gamma _{0}^{\prime }\approx -h_{1}\gamma _{0}^{3} + O(\gamma _{0}^{5}) ,
\label{68}
\end{equation}
which leads to
\begin{equation}
\gamma ^{\prime }\approx -h_{1}\gamma _{0}^{3}(1-h_{1}\gamma _{0}) +
O(\gamma_{0}^{5}) .   \label {69}
\end{equation}
The lesson we learn from the equation (\ref{66}) is that in all "correction"
terms
(which contain $h_{1}$, $h_{2} ...$) the expansion parameter $\gamma $ appears
in product with the rank $q\gamma $ which becomes large at high ranks, i.e. at
high multiplicities. Therefore for high multiplicity events one should take
into
account the ever higher order terms in $\gamma $. This problem was mentioned a
long time ago \cite{5} and discussed in some detail in \cite{38} but only
recently it has been analyzed.

As was mentioned, the double logarithmic formulae are obtained from
eq.(\ref{63})
for $h_{1}=h_{2}=0, \gamma =\gamma _{0}, \gamma ^{\prime }=0$. In that case
\begin{equation}
H_{q}\sim q^{-2} .  \label{70}
\end{equation}
It reminds the asymptotics of the negative binomial distribution with rather
small value of the parameter $k\approx 2$ and corresponds to the extremely wide
multiplicity distribution (see (\ref{23})) while experimental data provide
values
of $k$ ranging from 3.5 to $\sim $100.

Perhaps, the more interesting feature is the evolution of the qualitative
behaviour of the ratio $H_q$ at higher orders. In the modified leading
logarithm
approximation when $h_{1}$-term has been kept but $h_{2}$-term (as well as
higher order ones) neglected in (\ref{63}),
we observe that $H_{q}$ crosses the absciss axis, acquires a minimum at
\begin{equation}
q_{min}\approx \frac {1}{h_{1}\gamma _{0}} + \frac {1}{2}\approx 5   \label{71}
\end{equation}
and tends to zero from below as $\sim -q^{-1}$. If one includes the term with
$h_{2}$ too, the ratio $H_{q}$ gets the second zero and tends asymptotically to
a positive constant $h_{2}\gamma _{0}^{2}$. It reminds of the situation with
expansion of, e.g., $\cos x$ in Taylor series. That is why it does not
surprise us that one gets an oscillating behaviour of $H_q$ \cite{39} when
takes into account the higher order terms with $h_3$ and $h_{11}$ in the
formula (\ref{63}). The very first minimum reveals just the first oscillation.
It is demonstrated in Fig.6.  However, it has been shifted to $q\approx 4$ in
the approximation of \cite{39} what shows high sensitivity of $H_q$ to various
assumptions. Let us emphasize that the amplitudes of extrema and the
periodicity
of "quasi-oscillations" are different from all those shown in Figs.1 - 4.
The analogous behaviour of $H_q$ has been found as the exact solution of
QCD equations for fixed coupling constant (see section 7.2).

Thus we have demonstrated in this section that the conservation laws and other
higher order terms lead in the framework of gluodynamics to substantial
reduction of the width of the multiplicity distribution and to qualitative
change of the behaviour of cumulant and factorial moments.

\Section{Approximate solutions of QCD equations with running coupling constant}

Transition from gluodynamics to quantum chromodynamics where quarks are created
beside gluons leads back to the system of two coupled equations (\ref{50}),
(\ref{51}) for the generating functions of quarks and gluons instead of a
single equation (\ref{56}).
Their structure, however, does not differ, in principle, from the gluodynamics
equation described above in detail. That is why we shall not write down all the
relations (see, e.g., \cite{32, 40, 41}), and describe just the results
obtained.

In complete analogy to gluodynamics, one gets the system of the coupled
recurrent
equations for factorial moments and cumulants when the Taylor series expansion
is
used. It has been solved numerically \cite{40}. The properties of gluon jets do
not change noticeably, i.e. their cumulants and factorial moments are very
close
to those calculated in gluodynamics. Gluon ratio $H_q$ has the minimum at the
same value $q\approx $4 or 5. The quark factorial moments are larger than those
of gluon jets, i.e. parton multiplicity distribution for quark jets is wider
than
for gluon jets even though the average multiplicity is smaller there. First
minimum of quark cumulants and of their ratio to factorial moments is
positioned at $q\approx 8$.

To apply these results to the real process of the electron-positron
annihilation, one should relate its generating function to those for quark
and gluon jets. Having in mind the Feynman diagram of production of two quark
jets at the very initial stage one would write down
\begin{equation}
G_{e^{+}e^{-}}\approx G_{F}^{2} ,     \label{72}
\end{equation}
with further corrections (see, e.g., \cite{32}). In that case the zeros of
the quark jet cumulants and of $e^{+}e^{-}$ processes coincide because the
logarithms of generating functions which determine corresponding cumulants
(see (\ref{5})) are proportional to one another. It means that the first
minimum for $e^{+}e^{-}$ would lie at $q\geq 6$ since the first zero for quark
jets is positioned at $q>5$. The analysis of experimental data described
below (see Section 8) points out that this is not the case and either relation
(\ref{72}) should be revised or the higher order terms in Taylor series
expansion become crucial. The latter possibility appears less probable because
the similar shift of zeros of quark cumulants has been observed in case of
exact solution with the fixed coupling as described in the next section.
Independently of it, it seems that the most important
conclusion we get from the theoretical studies is the presence of maxima and
minima of the ratio $H_q$ which replace each other with some periodicity
(but not at neighbouring values of $q$ as it happens for fixed multiplicity).

It is interesting to note that the equations for the low order moments give
rise to conclusions about the anomalous dimension $\gamma $ and about the ratio
$r\equiv \langle n_{G}\rangle / \langle n_{F}\rangle $ of the average
multiplicities in gluon and quark jets \cite{41}. They have been represented
by the perturbative expansion as
\begin{equation}
\gamma = \gamma _{0}(1-a_{1}\gamma _{0}-a_{2}\gamma _{0}^{2}) , \label{X}
\end{equation}
\begin{equation}
r = \frac {N_c}{C_F}(1-r_{1}\gamma _{0}-r_{2}\gamma _{0}^{2}).  \label{Y}
\end{equation}
The coefficients $a_{i}, r_{i}$ have been calculated \cite{41}. They are shown
in Table 1 together with values of $r$ and $\gamma _{0}$ for various numbers of
flavours. Fig.7 demonstrates the $Q$-dependence of $\gamma $ resulting
from the solution of the equations in the higher order approximation discussed
above. The corresponding behaviour of the average multiplicity is shown in
Fig.8. For comparison, the energy ($y$) dependence of the mean multiplicity
for fixed coupling constant is drawn by dotted lines. At the very beginning it
increases rather slowly but at higher energies its asymptotical
increase exceeds that of the running coupling case. It is reliable since the
constant has been chosen at rather high energy (mass of $Z^0$) at $y_{Z^0}$=
6.67, i.e. its value is quite small. In real situation, it should increase
during the evolution of the jet, but the number of active flavours must
decrease. Let us note that these two trends somewhat compensate one
another in energy dependence of mean multiplicity. The ratio $r$ of the mean
multiplicities in gluon and quark jets is much smaller \cite{41} than its value
in the double logarithmic approximation where it is equal to 9/4. On the
average,
it is lower by about 20 per cents. The analogous situation is for exact
solutions
of equations with fixed coupling constant. Therefore we consider that ratio
in more detail in the next section.

\Section{Exact solutions of QCD equations with fixed coupling constant}

The above experience with QCD equations treated in various approximations
suggests that the conservation laws and non-singular terms of the
kernels play more important role than the dependence of the coupling constant
on the evolution parameter. It can be shown \cite{21}, \cite{42} that the
equations (\ref{50}), (\ref{51}) are solved exactly if the coupling constant is
fixed. No other assumptions are necessary. One gets the general solution for
the
moments of any rank but we start with the lowest ranks for pedagogical reasons.

\Subsection{First moments and the ratio of average multiplicities in gluon
to quark jets}

The equations for average multiplicities (unnormalized moments of first order)
are derived from the system of equations (\ref{50}), (\ref{51}) if one
substitutes
the generating functions as series (\ref{7}) and equates the linear in $z$
terms with the conditions $F_{0}=F_{1}=\Phi _{0}=\Phi _{1}=1$. (We denote the
factorial moments of quark jets by $\Phi _q $ and their cumulants by $\Psi
_q$.)
If the coupling constant is kept fixed, the average multiplicities behave
\cite{5} as
\begin{equation}
\langle n_{G}(y)\rangle = \e ^{\gamma y} ; \;\;\; \langle n_{F}(y)\rangle =
\e ^{\gamma y}/r ,   \label{73}
\end{equation}
where the anomalous dimension $\gamma $ and the ratio $r$ are constant. These
properties are inherited in equations (\ref{50}), (\ref{51}) as one notices
from the relations
\begin{equation}
\langle n_{G,F}(y+\ln x)\rangle /\langle n_{G,F}(y)\rangle = x^{\gamma } ,
\label{74}
\end{equation}
\begin{equation}
\langle n_{G,F}(y)\rangle ^{\prime } = \gamma \langle n_{G,F}\rangle .
\label{75}
\end{equation}
Then the equations  are rewritten as a system of two algebraic equations for
two variables $\gamma $ and $r$:
\begin{equation}
\gamma = \gamma _{0}^{2} [M_{1}^{G} + n_{f}r(M_{1}^{F} - M_{0}^{F})] ,
\label{76}
\end{equation}
\begin{equation}
\gamma = \gamma _{0}^{2} [L_{2} - L_{0} + rL_{1}] ,  \label{77}
\end{equation}
where
\begin{eqnarray*}
M_{1}^{G} = \int _{0}^{1} dxK_{G}^{G}[x^{\gamma }+(1-x)^{\gamma }-1] , \\
M_{1}^{F} = \int _{0}^{1} dxK_{G}^{F}[x^{\gamma }+(1-x)^{\gamma }] ,  \\
M_{0}^{F} = \int _{0}^{1} dxK_{G}^{F} = M_{1}^{F}(\gamma =0)/2 ,  \\
L_{1} = \int _{0}^{1} dxK_{F}^{G}x^{\gamma } ,   \\
L_{2} = \int _{0}^{1} dxK_{F}^{G}(1-x)^{\gamma } ,  \\
L_{0} = \int _{0}^{1} dxK_{F}^{G} = L_{1}(\gamma =0) . \\
\end{eqnarray*}
Coefficients $M_{i}, L_{i}$  can be expressed in terms of Euler beta-functions
and psi-functions and depend on $\gamma $ only. At fixed $\gamma _{0}$, both
$\gamma $ and $r$ are constant. Let us stress that $\gamma $ is not equal to
$\gamma _{0}$ even in gluodynamics at $n_{f}=0$ because $M_{1}^{G}$ differs
from $\gamma ^{-1}$. The approximate equality is valid for $\gamma _{0}\ll 1$
but the perturbative expansion for $\gamma $ differs from the corresponding
formula (\ref{67}) for the running coupling constant
\begin{equation}
\gamma \approx \gamma _{0} - h_{1}\gamma _{0}^{2} + \frac
{1}{2}(h_{1}^{2}+h_{2})
\gamma _{0}^{3}+O(\gamma _{0}^{4}) ,        \label{78}
\end{equation}
wherefrom one sees that the first correction is twice larger.

The ratio $r$ appears in equations (\ref{78}), (\ref{79})linearly, and one
is tempted to rewrite them as
\begin{equation}
r(\gamma ) = b(\gamma )/ \left [\frac {\gamma }{\gamma _{0}^{2}} - a(\gamma )
\right ] ,    \label{79}
\end{equation}
\begin{equation}
r(\gamma ) = \left [\frac {\gamma }{\gamma _{0}^{2}} - d(\gamma )\right ]
/c(\gamma ) ,  \label{80}
\end{equation}
where
\begin{eqnarray*}
a = \psi (1)-\psi (\gamma +1)+B(\gamma ,1)-2B(\gamma +1,2)-2B(\gamma +2,1)+
B(\gamma +2,3)+B(\gamma +3,2)+11/12-n_{f}/6N_{c} ,   \\
%\end{equation}
%\begin{equation}
b = \frac{n_{f}}{2N_{c}}[B(\gamma +3,1)+B(\gamma +1,3)] ,  \\
%\end{equation}
%\begin{equation}
c = \frac {C_{F}}{N_{c}}[B(\gamma ,1)-B(\gamma +1,1)+B(\gamma +2,1)/2] ,  \\
%\end{equation}
%\begin{equation}
d = \frac {C_{F}}{N_{c}}[\psi (1)-\psi (\gamma +1)-B(\gamma +1,1)+
B(\gamma +1,2)/2 +3/4] .   \\
\end{eqnarray*}
All beta-functions are just the inverse polynomials of $\gamma $ but the
above notations are less cumbersome. The solution of the algebraic relations
(\ref{79}), (\ref{80}) provides $\gamma $ and $r$ as functions of $gamma _{0}$
and $n_{f}$. Fig.9 shows the dependence of $\gamma $ on $\gamma _{0}$ for
$n_{f} $ = 3, 4, 5. The differences for the various values of $n_{f}$ are
hardly
discernable, being less than the thickness of the visible line. Note that
$\gamma $ is significantly different from $\gamma _{0}$. They can be related
by the simple fitted formula
\begin{equation}
\gamma  = 0.077 + 0.62\gamma _{0}      \label{81}
\end{equation}
or the more theoretically motivated one (\ref{78}) starting from the linear
terms
in $\gamma _{0}$
\begin{equation}
\gamma = 0.97\gamma _{0} - 0.48\gamma _{0}^{2} + 0.2\gamma _{0}^{3}
\label{82}
\end{equation}
fitted by computer in the range of $\gamma _{0}$ from 0.48 to 0.6. Let us note
that $\gamma $ changes very slowly with $\gamma _{0}$. By itself, the value
of $\gamma $ is not of much interest even though it is related to the energy
dependence of the average multiplicity. However, it is known that the power
increase of the mean multiplicity for fixed coupling is replaced by the
dependence $\sim \exp (\ln s)^{1/2}$ for running coupling. Somehow the reduced
value of $\gamma $ compared to $\gamma _{0}$ respects that tendency but the
dependence (\ref{73}) can not be used in asymptotics. More realistic behaviour
provided by the running coupling has been discussed in the previous section
(see Fig.8).

The corresponding ratio $r$ is of more interest since the energy dependences
of the average multiplicities in gluon and quark jets cancel. That is why its
prediction for fixed coupling could be  more general. The corresponding result
on the ratio $r$ is shown in Fig.10. Again the dependence on $n_{f}$ is very
mild,
and is exhibited in the expanded scale in Fig.10. More important, the
dependence
of $r$ on $\gamma _{0}$ is even weaker than that of $\gamma $, and the
average effective value is equal to
\begin{equation}
r = 1.84 \pm 0.02 .    \label{83}
\end{equation}
Such a low value of the ratio $r$ should arise interest since in the double
logarithmic approximation it is much larger \cite{5}, and equal to
$N_{c}/C_{F}$
=9/4. It has been reduced to 2.05 in the modified  leading logarithm
approximation
\cite{43},\cite{44}. The above value shows that the conservation laws diminish
the ratio further. Even somewhat lower values of $r$ have been obtained for
running coupling (see \cite{41} and Table 1).

In a realistic process, the virtuality in a jet degrades as partons evolve
toward hadrons, presumably with the associated change of the number of active
quark flavours. In the framework of our calculations with fixed coupling, this
dependence could be considered using the formula
\begin{equation}
\gamma _{0}^{2} = \frac {12}{\beta _{0}y} , \;\;\;\;\; \beta _{0} =
11-\frac {2n_{f}}{3}  \label{84}
\end{equation}
for $y = \ln Q/Q_{0} , Q_{0} = 0.65/\Lambda _{\overline {MS}}$, where we
include
the dependence of $\Lambda _{\overline {MS}}$ on $n_f$ according to the
proportions 63 : 100 : 130 for $n_{f} = 5 : 4 : 3$, respectively \cite{45}, and
we pin the
value $\Lambda _{\overline {MS}}$ at 175 MeV for $n_{f} = 5$. We consider the
values of $Q$ at $M_{Z}/m$ ($M_{Z}$ being the mass of $Z^{0}$) for $m$ = 1, 2,
4,
and 8. The result on $\gamma $ is shown as a function of $\ln Q$ in Fig.11.
The dependence on $n_f$ is so small that connection of the different points
with
different $n_f$ for the same $Q$ results only in short line segments as shown.
Thus we learn that as a jet of partons evolves toward lower $Q$, we need not be
concerned with the change of the active flavour, and that the parton
multiplicity will depend on the evolving virtuality through a mild variation of
$\gamma $, but not enough to invalidate fixed coupling approximation.
Certainly,
in the ratio of the multiplicities, such dependences are cancelled, yielding
a stable value of $r$. This conclusion has been supported by the approximate
solutions of the equations with the running coupling \cite{41} as we have
discussed already (see Table 1).

\Subsection{Higher order moments and widths of distributions in gluon and quark
jets}

The dispersion of the multiplicity distribution is determined by the second
moment. Therefore, to get it, one should solve the system of equations
(\ref{50}),
(\ref{51}) for $q = 2$. However, the relations (\ref{74}), (\ref{75}) prompt us
to get the solution for any $q$. In fact, one can obtain the system of coupled
recurrent equations \cite{21} for moments if one substitutes in equations
(\ref{50}), (\ref{51}) the generating functions according to (\ref{7}) and
compares the coefficients in front of $z^q$ in both sides. Those equations are
solved by iteration. It is well suited for computer calculations. We do not
write them down here (see \cite{21}), and show just the
final analytical expressions for the moments of the rank $q$ as related to the
lower rank moments. For that purpose, let us introduce
\begin{equation}
f_{q} = F_{q}/q! ,  \;\;\;\;\;  \hat{\phi }_{q} = \Phi _{q}/r^{q}q! .
\label{84}
\end{equation}
The solution of the equation is \cite{21}
\begin{equation}
f_{q} = [a_{q}S_{q}(f,\hat{\phi }) + b_{q}T_{q}(f,\hat{\phi })]/\Delta _{q} ,
\label{85}
\end{equation}
\begin{equation}
\hat{\phi }_{q} = [c_{q}S_{q}(f,\hat{\phi }) + d_{q}T_{q}(f,\hat{\phi
})]/\Delta _{q} ,
\label{86}
\end{equation}
where
\begin{equation}
S_{q} = \sum _{l=1}^{q-1}[N_{q,l}^{G}f_{l}f_{q-l} + n_{f}N_{q,l}^{F}
\hat{\phi }_{l}\hat{\phi }_{q-l}] ,  \label{87}
\end{equation}
\begin{equation}
T_{q} = \sum _{l=1}^{q-1}L_{q,l}\hat{\phi }_{l}f_{q-l} ,   \label{88}
\end{equation}
\begin{equation}
a_{q} = \frac {q\gamma }{\gamma _{0}^{2}} + L_{0,0} - L_{q,q} ,  \label{89}
\end{equation}
\begin{equation}
b_{q} = n_{f}M_{q}^{F} ,   \label{90}
\end{equation}
\begin{equation}
c_{q} = L_{q,0} ,  \label{91}
\end{equation}
\begin{equation}
d_{q} = \frac {q\gamma }{\gamma _{0}^{2}} - M_{q}^{G} + n_{f}N_{0,0}^{F} ,
\label{92}
\end{equation}
\begin{equation}
\Delta _{q} = a_{q}d_{q} - b_{q}c_{q}  ,  \label{93}
\end{equation}
\begin{eqnarray*}
M_{q}^{G} = \psi (1) - \psi (q\gamma +1) + B(q\gamma ,1) - 2B(q\gamma +1,2) -
2B(q\gamma +2,1) + B(q\gamma +2,3) + B(q\gamma +3,2) +11/12 ,  \\
%\end{equation}
%\begin{equation}
M_{q}^{F} = \frac {1}{N_{c}}[B(q\gamma +3,1) + B(q\gamma +1,3)] , \\
%\end{equation}
%\begin{equation}
N_{q,l}^{G} = B(l\gamma ,(q-l)\gamma +1) - 2B(l\gamma +1,(q-l)\gamma +2) +
B(l\gamma +2,(q-l)\gamma +3) ,    \\
%\end{equation}
%\begin{equation}
N_{q,l}^{F} = \frac {1}{4N_{c}}[B(l\gamma +3,(q-l)\gamma +1) + B(l\gamma +1,
(q-l)\gamma +3)] , \\
%\end{equation}
%\begin{equation}
L_{q,l} = \frac {C_{F}}{N_{c}}[B(l\gamma +1,(q-l)\gamma ) - B(l\gamma +1,
(q-l)\gamma +1) + B(l\gamma +1,(q-l)\gamma +2)] .  \\
\end{eqnarray*}
The above expressions look cumbersome {\footnote The gluodynamics formulae
are obtained from them in the limit $n_f = C_F = 0$ when leading terms of
$M_{q}^{G}$ and of $N_{q,l}^{G}$ are considered only, i.e. $B(q\gamma ,1)\equiv
1/q\gamma$ and $B(l\gamma ,(q-l)\gamma + 1)$. }
but their structure is very simple and
clear. They generalize the formulae of the preceding section to any $q$. The
formulae of gluodynamics follow from them for $n_{f} = C_{F} = 0$ if one leaves
in
$M_{q}^{G}$ and $N_{q,l}^{G}$ the leading terms $B(q\gamma ,1)\equiv 1/q\gamma
$
and $B(l\gamma ,(q-l)\gamma +1)$. Using the values of $\gamma $ and $r$ from
the
preceding section at given $\gamma _{0}$ and $n_{f}$, one gets first $F_2$ and
$\Phi _2$, and then increases $q$ by 1.

The evolution parameter $y$ disappears from the formulae. {\it A'posteriori},
it
means that our assumption about $F$--scaling with all dependence on $y$ hidden
in
average multiplicities $\langle n_{G,F}\rangle (y)$ is correct for fixed
coupling.
It leads to the self-consistent system of algebraic equations where all
quantities, including $F_{q}$ and $\Phi _q$, do not depend on energy.
Surely, $F$--scaling is precise at fixed coupling until the main equations
are correct. Actually, one should speak about the asymptotic $F$--scaling
because the limits of $x$-integration in eqs. (\ref{50}), (\ref{51}) are
asymptotic ones. More precise treatment of them would correspond to considering
the higher twist effects.

The results of calculation, when expressed in terms of $F_q$ and $\Phi _q$,
are shown in Fig.12  for $\gamma _{0} = 0.48$ and $n_{f} = 5$. Evidently, they
increase rapidly with $q$, more so for $\Phi _q$ than $F_q$. Since these are
normalized factorial moments, they imply that the multiplicity distribution
for the quark jet is wider than that for the gluon jet, although the average
multiplicity in quark jets is lower than in gluon jets. These results are
very insensitive to the number of active flavours $n_{f}$. The dependence
on the coupling constant is very mild and the results are rather insensitive to
coupling constant being fixed or running.

Let us compare QCD results to those of the phenomenological distributions. By
comparing Fig.12 with Fig.2 and Fig.3, the QCD results are clearly of the NBD
type
rather than  the fixed multiplicity. In fact, $F_q$ in Fig.12 can be
approximated
by the negative binomial distribution with $k=5$. However, it is an apparent
coincidence. While the characterization of $F_q$ by the NBD parameter $k$ is
convenient, the fits by the negative binomial distribution are inappropriate.
Let us recall that the cumulants of the negative binomial distribution decrease
at low $q$ and then increase. The ratio $H_q$ decreases monotonically with $q$.

We define the corresponding ratios for gluon and quark jets as:
\begin{equation}
H_{q} = K_{q}/F_{q} ,  \label{94}
\end{equation}
\begin{equation}
\eta _{q} = \Psi _{q}/\Phi _{q} , \label{95}
\end{equation}
where $K_{q} (\Psi _{q})$ are related to $F_{q} (\Phi _{q})$ by the formula
(\ref{11}). The results of calculations in the fixed coupling QCD are shown in
Fig.13 and Fig.14 .

The distinctive feature of the behaviour of $H_{q}$ is clearly its
oscillations.
There are no oscillations for the negative binomial distribution even if it
fits
the second and third moments quite well. One sees that the fixed multiplicity
distribution, which gives rise to oscillations of $H_q$ changing sign at each
subsequent integer value of $q$, does not suit us because of wrong values of
moments and improper period of oscillations (see the discussion of experimental
results in the following section).

The sensitivity of $H_q$ to the shapes of the distributions obtained in quantum
chromodynamics at different assumptions is clearly demonstrated by its various
qualitative forms. It is positive and decreases monotonically as $\sim q^{-2}$
in the double logarithmic approximation, acquires a zero and a minimum in the
modified leading logarithm approximation tending asymptotically to zero from
below
like $\sim -q^{-1}$, gets the second zero and constant positive asymptotics
due to the next term and starts oscillating in the higher order approximations.

The behaviour of $H_q$ depends strongly on slight variations of the particular
shape of factorial moments at low values of $q$. One can demonstrate how easy
it is to get the oscillations of the fixed multiplicity type imposed on the
double logarithmic behaviour by doing the following exercise. It is known
\cite{19, 46} that the large $q$ behaviour of $F_q$ of the factorial moments in
the
double logarithmic approximation (see (\ref{23})) is
\begin{equation}
F_{q} = \frac {2q\Gamma (q+1)}{C^{q}} .    \label{96}
\end{equation}
If we add a preasymptotic term by replacing the factor $2q$ by $2q+1$ in the
numerator (it restores the condition $F_{0}=1$ but not $F_{1}=1$), one gets an
additional term in $H_{q}$ which imposes the oscillations of the fixed
multiplicity type on the monotonous decrease of the $q^{-2}$ shape, and the
ratio $H_q$ becomes
\begin{equation}
H_{q} = \frac {2+(-1)^{q-1}}{q(2q+1)} ,  \label{97}
\end{equation}
where the second term in the numerator appears due to the newly added
preasymptotic term.

The above examples show how sensitive $H_q$ is to various approximations made
in solving the set of equations for generating functions. Their distinction has
been demonstrated in Figs.4, 6, 13, 14, and they strongly differ from the
phenomenological distributions shown in Figs.2, 3. Unfortunately, there
is still no clear understanding of the physical origin of oscillations, i.e. of
their
periodicity, amplitude and their dependence on the rank $q$ (it seems that the
amplitude increases and the period decreases with $q$). Nevertheless, the exact
solution of fixed coupling QCD equations provides clear guide to the realistic
behaviour of $H_q$. Perhaps, the behaviour of $H_q$ will show us the ways to
generalize the equations for the generating functions including the fine
effects
of the interaction of "colour monsters" \cite{5, 19, 46}.

Let us point out that the above oscillations proceed at integer values of $q$
and are not related to the fractional moments oscillations. The latter ones
would impose the lower period harmonics on those oscillations.

\Section{Experiment}

Thus we have obtained the results for the KNO--function $f(x)$, for the moments
of the multiplicity distribution $K_{q}, F_{q}$ and for their ratio $H_{q}$ as
well as for the energy behaviour of mean multiplicities (the anomalous QCD
dimension $\gamma $) and for the ratio $r$ of the average multiplicities in
gluon and  quark jets. Before
comparing them to experiment one should remind that all above results have been
obtained for the multiplicity distributions of partons (gluons and quarks)
while
experimentators have to deal with hadrons. To translate theoretical predictions
to experimentally measured values, one should elaborate a hadronization model
describing the transition from partons to hadrons. After that, one can get the
quantitative results using Monte Carlo calculations. Sometimes, one relies on
the
hypothesis of the local parton-hadron duality. It assumes that the
distributions
of partons and hadrons differ by the numerical coefficient only which is
determined
by the number of partons recombined in a single hadron. Therefore, it is
unimportant for the normalized variables, and the normalized moments of gluon
and
quark jets should be just related to the moments of observed processes, e.g.,
to electron-positron annihilation. On the contrary, such values as the average
multiplicities in gluon and quark jets are changed in a different way what can
vary their ratio also. In such a case one should use some Monte Carlo versions
of hadronization. One of them, named Herwig \cite{47}, has been discussed in
\cite{48} in relation to the ratio of average multiplicities $r$ (\ref{73})
both
on parton and hadron levels. Its results are shown in Fig.15.
It supports the asymptotical local parton-hadron
duality but does not respect it at intermediate energies in exact way.
The partonic ratio is equal to $r_{parton}^{MC}\approx 1.9$. It shows
that the higher order corrections play an important role in Herwig because this
value is smaller than those of the lowest order approximations as discussed
above.
The hadronic ratio is slightly lower in asymptotics. However, it is still much
lower at energies of $Z^{0}$ (equal to 1.44) and increases with energy. It
shows
that the local parton-hadron duality is not yet very accurate. The model
describes the bulk of experimental data even though it admits slight revisions.
In particular, if one modifies the partonic cascade so that it gives rise to
the
value of the partonic ratio obtained above $r_{theor}=1.84\pm 0.02$ (\ref{83})
and considers the same hadronization model to deal with the same ratio of
hadrons to partons $r_{hadron}^{MC}/r_{parton}^{MC}=1.44/1.92$ (see Fig.15),
i.e. applies the formula
\begin{equation}
r_{exp} = r_{theor}(r_{hadron}^{MC}/r_{parton}^{MC}) ,  \label{98}
\end{equation}
one obtains
\begin{equation}
r_{exp} \approx 1.38\pm 0.02 .  \label{99}
\end{equation}
It agrees quite well with the recently measured \cite{49} value $1.27\pm 0.04
\pm 0.06$ (see also \cite{50}). Therefore we can state that there is no
disagreement between theoretical
and experimental values of the ratio of average multiplicities in gluon  and
quark jets any more. Still one should keep in mind that the phenomenologically
described hadronization plays an important role reducing this value by 40
per cents when going from partonic to hadronic levels.

What concerns the KNO--function $f(x)$, it becomes narrower when the energy
conservation is taken into account in a proper way compared to the double
logarithmic approximation as was discussed in Section 5.1. The negative
binomial distribution fits it rather well with $k\approx 7$ as is seen from
Fig.5. Monte Carlo calculations fit experiment as well. However we would
like to stress once again that some tiny features escape such a comparison,
and they can be revealed when studying the behaviour of the ratio of cumulant
to factorial moments $H_q$.

These ratios differ for gluon and quark jets as shown in Figs.13, 14.
Therefore,
they may be used either for selection of those jets, or for the control of
validity of their separation done according to other criteria (see, e.g.,
\cite{49},\cite{50}). The qualitative characteristics of the curves are
hardly changed by hadronization even though one should confirm it by Monte
Carlo
calculations that has not been done yet.

The way from the generating functions of jets to real processes of
hadroproduction
is non-trivial one even for $e^{+}e^{-}$-annihilation as has been discussed
above,
and ever so more complicated for other processes. One can hope that the
qualitative
features of jet moments are more general. Then it is reasonable to confront
them to the corresponding characteristics of multiparticle production processes
in attempt to reveal their interrelation and, in particular, the role of
hadronization.

Such an analysis has been performed \cite{51} both for
$e^{+}e^{-}$-annihilation
and for $pp$ (and $p\bar{p}$)-interactions in a wide energy interval to get
some guidance to the difference between the processes initiated by leptons and
by hadrons. A list of the considered samples of experimental data \cite{52}--
\cite{65} is given in
Table 2. The low statistics samples, yielding large error bars, have been
disregarded, and only papers reporting a detailed separation between elastic
and
inelastic data for low multiplicities have been taken into account. For all the
considered experimental multiplicity distributions of secondary hadrons
the ratio $H_q$ has been computed up to the 16-th order. It happens that the
qualitative features of the behaviour of $H_q$ are very similar in all
processes,
though the detailed structure depends on the type of the interaction, on
primary
energy and even on the experimental selection of events.

As a first example one may consider the outcomes of the $e^{+}e^{-}$ 91 GeV
DELPHI multiplicity distribution \cite{55} plotted in Fig.16. Due to different
order
of magnitude involved here, the low-$q$ and high-$q$ regions are plotted using
different scales. The polyline in the upper right box displays, up to the
fourth
order, the abrupt descent of $H_q$ (the scale is logarithmic). In the main
reference frame two negative minima (at $q\approx $ 5 and 12) and two positive
maxima (at $q\approx $ 8 and 15) are shown. The predictions of the negative
binomial distribution are given by the dashed curve shown in Fig.16. The value
of the parameter $k$ has been taken from ref.\cite{55} ($k^{-1}=0.0411\pm
0.0012$). No minima or maxima appear, of course, and one is tempted to conclude
that the negative binomial distribution is able to reproduce the main details
of
experimental data at low $q$ but not its tiny details revealed by oscillations
at higher $q$.

Fig.17 refers to the $\overline{p}p$ 546 GeV interaction outcomes from the UA5
collaboration \cite{65}. The main difference with respect to the previous case
is the order of magnitude of maxima and minima, whose absolute value is more
than ten times larger. Nonetheless, the same main characteristics found for
$e{+}e^{-}$-annihilation, i.e. the abrupt descent and the subsequent
oscillations, can be observed (here  minima are at $q\approx $ 6 and 12 while
maxima are at $q\approx $ 9 and 15). Again superimposed (dashed curve) are
the negative bimomial yields, here corresponding to the $k$ parameter given
in ref.\cite{65} ($k=3.69\pm 0.09$).

The similar {\it qualitative} peculiarities have been observed at various
energies as shown in Fig.18 and Fig.19. Let us point out that there is
{\it quantitative}  distinction between the experimental data of different
collaborations for the {\it same} $e^{+}e^{-}$-process at the {\it same} energy
91 GeV (see the four low graphs in Fig.18). Perhaps, it is related to the
selection procedures adopted by various collaborations and to systematical
errors. The high sensitivity of the ratio $H_q$ could be exploited, e.g., for
optimization of selection criteria. To check their influence, the various
Monte Carlo schemes have been confronted to the data about $H_q$. It has
happened that ARIADNE which includes the selection criteria of OPAL describes
its data in Fig.18g while JETSET suited for DELPHI fits its data in Fig.18f.
Since there is
quantitative difference between those data, one concludes that it stems from
the particular details of experiments but not from the general dynamics of the
process at the parton level. The general trend survives various systematics.

Let us note that the distinction between experimental and negative binomial
distributions has been noticed in papers of UA5 \cite{65}, DELPHI \cite{55}
and OPAL \cite{57}. The latter collaboration has shown that the difference
between the two distributions oscillates. Perhaps it lies at the origin of the
oscillations of $H_q$ too. Their physical interpretation could be related to
the
varying number of subjets in $e^{+}e^{-}$-annihilation (see, e.g., \cite{31})
or to the number of "ladders-strings" in hadronic processes (see, e.g.,
\cite{27}).

The more trivial effect of the cut-off of the experimental distributions
at large multiplicities could be of importance since it produces some
oscillations of $H_q$ also. In distinction to QCD, this effect should,
however, vanish at asymptotically high energies. Nevertheless, it requires
further study to be done.

To conclude, we say that the quantum chromodynamics predictions became more
reliable at the qualitative level during the last years and got support from
experimental data on multiplicity distributions. There exist now both the
proper
approach to treatment of various theoretical approximations and some proposals
of the selection of experimental data for quantitative comparison with theory
predictions. Therefore the solid foundation has been laid for the precise
description of the problem as a whole.

\Section{Evolution of distributions with decreasing phase space --
intermittency
and fractality}

The multiplicity distributions can be measured not only in the total phase
space (as it has been discussed above for very large phase space volumes) but
in any part of it. For the homogeneous distribution of particles within the
volume, the average multiplicity is proportional to the volume and decreases
for small volumes but the fluctuations increase. The most interesting problem
here is a regularity in the increase of fluctuations, the probable
distinction from the purely statistical laws related to the decrease of the
average multiplicity. Such a variation has to be connected with the dynamics
of interactions. In particular, it has been proposed \cite{66} to look for the
power-law behaviour of the factorial moments for small rapidity intervals
$\delta y$
\begin{equation}
F_{q}\sim (\delta y)^{-\phi (q)} ,  \;\;\;\;\;\; (\delta y \rightarrow 0) ,
\label{100}
\end{equation}
where $\phi (q)>0$. This assumption has been provoked by the analogy to the
turbulence in hydrodynamics, where the similar property is known as
intermittency and $\phi (q)$ are called the intermittency exponents. It
originates from the state of the fluid where the "quiet" regions alternate with
the volumes of high fluctuations, and it becomes even more noticeable at
smaller sizes. From the point of view of distributions it implies rather strong
increase of their width with slower decline at high multiplicities.

Experimental data on various processes in the wide energy range support the
idea
by revealing the power-law dependence (\ref{100}). Immediately, several
theoretical approaches were developed to explain that phenomenon. The present
state of affairs has been described recently in the review paper \cite{14}.
Here
we just show how quantum chromodynamics reproduces intermittency
\cite{38,67-71}.

Let us stress once again that quantum chromodynamics deals with partons (quarks
and gluons) while experimental results provide the moments of distributions of
hadrons. The local parton-hadron duality hypothesis implies proportionality of
inclusive distributions but is not so obvious for correlations and is not
fulfilled sometimes in the proposed Monte Carlo schemes. Therefore, we can
pretend to get qualitative description on the parton level but not to attempt
quantitative comparison to experiment.

In contrast to the previous sections, we rely here on the diagrammatic
approach instead of the equations for the generating functions. It is caused
by the fact that, considering the multiplicity distributions in small phase
space
volumes, one has to deal with a minor part of the whole parton content of a
well
developed jet, namely, with those partons which fill in the volume chosen.
Surely, the prehistory of a jet as a whole is important for the subjet under
consideration as is shown in Fig.20. Here    \\
\begin{enumerate}
\item the primary quark (solid line) emits the hard gluon with energy $E$ in
the direction of the angular interval $\theta $, but not necessarily hitting
the window,                                    \\
\item the emitted gluon produces the jet of partons with parton splitting
angles {\it larger} than the window size,        \\
\item among those partons there exists such a parton with energy $k$
which hits the window,                             \\
\item all decay products of the subjet cover exactly the bin $\theta $. \\
\end{enumerate}
This picture dictates the rules of calculation of the $q$-th correlator of the
whole jet. One should average the $q$-th correlator of the subjet $\Delta N
^{(q)}(k\theta }$ over all possible ways of its production i.e. convolute it
with the inclusive spectra of such partons $D^{\theta }$ in the whole jet  and
with the probability of creation of the jet ($\alpha _{S}K_{F}^{G}$).
Analytically, it is represented by
\begin{equation}
\Delta N^{(q)}\left ( Q\theta _{0}, \frac {\theta _{0}}{\theta }\right )
\propto
\int ^{Q} \frac {dE}{E} \frac {\alpha _{S}}{2\pi }K_{F}^{G}\left (\frac {E}{Q}
\right )\int ^{E}\frac {dk}{k}D^{\theta }\left (\frac {E}{k};E\theta _{0},
k\theta \right )\Delta N^{(q)}(k\theta ) ,  \label{101}
\end{equation}
where $\Delta N^{(q)}\equiv F_{q}\langle n \rangle ^{q}$ is the unnormalized
factorial moment (in the left-hand side, for the whole jet, and in the
right-hand
side, for the parton subjet with momentum $k$ hitting the bin $\theta $). Since
the unnormalized moments increase with energy while the parton spectrum
decreases,
the product $D^{\theta }\Delta N^{(q)}(k\theta )$ has a maximum at some energy,
and the integral over momenta may be calculated by the steepest descent method
. Leaving aside the details of calculations (see \cite{38}), we describe
the general structure of the correlator for the fixed coupling constant
$\gamma _{0} = \const $:
\begin{equation}
\Delta N^{(q)} \propto \Delta \Omega \left ( \frac {\theta _{0}}{\theta }\right
)
^{\frac {\gamma _{0}}{q}}\left ( \frac {E\theta }{c }\right )^{q\gamma _{0}} ,
\;\;\;\;\;\;  (c = \const ) ,   \label{102}
\end{equation}
where the three factors represent the phase space, the energy spectrum factor
and
the $q$-th power of the average multiplicity. To get the normalized moment, one
should divide (\ref{102}) by the $q$-th power of that part of the mean
multiplicity
of the whole jet which  appears inside the window $\theta $ i.e. by the share
of the total average multiplicity corresponding to the phase space volume
$\Delta \Omega $:
\begin{equation}
\Delta N(\theta ) \sim \Delta \Omega \Delta N(\theta _{0}) .  \label{103}
\end{equation}
If the analysis has been done in the $D$-dimensional space, the phase space
volume is proportional to:
\begin{equation}
\Delta \Omega \sim \theta ^{D} ,  \label{104}
\end{equation}
where $\theta $ corresponds to the minimal linear size on the $D$-dimensional
window that stems from the singular behaviour of parton propagators in
quantum chromodynamics (see \cite{38}). That is why the factorial moments may
be represented as products of the purely kinematical factor depending on the
dimension of the analyzed space, and of the dynamical factor, which is not
related to the dimension and defined by the coupling constant
\begin{equation}
F_{q} \sim \theta ^{-D(q-1)}\theta ^{\frac {q^{2}-1}{q}\gamma _{0}} .
\label{105}
\end{equation}
At small angular windows one gets $\theta \sim \delta y$ and the intermittency
indices defined by eq. (\ref{100}) are
\begin{equation}
\phi (q) = D(q-1) - \frac {q^{2}-1}{q}\gamma _{0} .  \label{106}
\end{equation}
The formula (\ref{106}) is valid for moderately small bins when the condition
$\alpha _{S}\ln \theta _{0}/\theta <1$ is fulfilled. For extremely small
windows, one should take into account that the coupling constant is running.
Then the constant $\gamma _{0}$ should be replaced by the effective value
$\langle \gamma \rangle $, which depends logarithmically on the width of the
window $\theta $ and may be approximated by \cite{38}
\begin{equation}
\langle \gamma \rangle = \gamma _{0}(1+\frac {\epsilon }{4}) , \label{107}
\end{equation}
where
\begin{equation}
\epsilon = \frac {q^{2}+1}{q^{2}}\frac {\ln \theta _{0}/\theta }{\ln (E\theta
_{0}/\mu } \leq 1 .  \label{108}
\end{equation}
As a result, numerical values of the intermittency indices for very small bins
become noticeably smaller than in the fixed coupling regime, especially for
the low-rank moments. Moreover, the simple power-law behaviour (\ref{100})
becomes
modified by the logarithmic correction terms and the intermittency indices
depend on the value of the interval chosen. The resulting curve of $\ln F_{q}$
as a function of $-\ln \theta $ has two branches. The rather steep linear
increase at the moderately small windows $\theta $ with the slope (\ref{106})
is
replaced at smaller bin sizes by much slower quasi-linear increase given by
(\ref{107}), (\ref{108}). It is easy to calculate the position of the
transition
point to another regime and show that at higher values of $q$ that point shifts
to smaller bin sizes. Still, the factorial moments of any rank increase at
smaller
intervals. It demonstrates that the corresponding fluctuations of the
multiplicity distributions become stronger than those in larger
intervals and, what is important, exceed noticeably the Poisson fluctuations.

We have described the results of the double logarithmic approximation. The
corrections due to the modified leading logarithm terms are comparatively
small numerically (about 10 per cents). For example, they move the transition
point (from power-law to quasipower regime) to slightly smaller bins for any
moment except of the second one where it moves to somewhat larger angles.
This tendency can be easily prescribed to the mutual influence of the energy
spectrum and  the average multiplicity. Of more importance are the qualitative
effects of new functional dependence on the rank $q$ due to the terms
proportional to $q\gamma $. For example, the attempt to exploit the analogy to
statistical mechanics \cite{38}, where the quantity $1-\frac {\phi (q)+1}{q}$
is interpreted as "free energy" and the rank $q$ as an inverse temperature
$\beta =1/kT$, has led to the unexpected result about the "phase transition".
It shows up because the "free energy" increases monotonically with $q$ in the
lowest order approximation while higher order corrections give rise to the
maximum just at those $q$ values where $H_{q}$ has a minimum (\ref{71})
described above i.e. at
\begin{equation}
q_{cr} \approx \frac {1}{h_{1}\gamma _{0} \approx 5 .   \label{109}
\end{equation}
In statistical mechanics, it would correspond to zero entropy i.e.
to the phase transition. Here, it just indicates the role of the new parameter
$q\gamma $ in quantum chromodynamics as it is discussed above.

The above results may be restated in terms of fractals. The power-like
behaviour of factorial moments points out to fractal properties of particle
(parton) distributions in the phase space. According to the general theory
of fractals (see \cite{14} and references therein), the intermittency indices
are related to fractal (Renyi) dimensions $D_{q}$ by the formula
\begin{equation}
\phi (q) = (q-1)(D-D_{q}) ,  \label{110}
\end{equation}
wherefrom one gets in the double logarithmic approximation taking into account
(\ref{106})
\begin{equation}
D_{q} = \frac {q+1}{q}\gamma _{0} = \gamma _{0} + \frac {\gamma _{0}}{q} .
\label{111}
\end{equation}
The first term corresponds to monofractal behaviour and is due to the average
multiplicity increase. The second one provides multifractal properties and is
related to the descent of the energy spectrum as discussed above. One can
easily get the multifractal spectral function in that case (see \cite{38}). It
is clearly seen that the fractality in quantum chromodynamics has a purely
dynamical origin ($D_{q}\sim \gamma _{0}$) related to the cascade nature of the
process while the kinematical factor in (\ref{105}) has an integer dimension.

The fractality of the particle distributions in the phase space volume could
suggest the fractal properties of colliding objects in the ordinary space-time.
Surely, due to its dynamical origin, such a structure would be dynamical
itself, i.e. fastly evolving in space and time. There are two reasons to
favour this possibility. First, the cascade process of the evolution of the
parton shower in the ordinary space must give rise to the "tree-like"
structure of the fractal type which should evolve due to the cascade evolution.
Second, the lattice computation in $SU(2)$-gluodynamics \cite{72} has shown
that
the system of gluons in the very vicinity of the phase transition is fractal
in the sense that it fills in the volume $V$ embordered by the surface $S$
which are related by the formula $V\sim S^{1.12}$. It is typical for fractal
objects since the exponent should be equal to 1.5 for ordinary
three-dimensional
objects.

The geometrical fractality of the macroscopic bodies has been revealed by
measurements of the power-like shape of the structure functions when
some point-like particles (photons, electrons, neutrons etc) are scattered
by them. Using this example, one could try to propose \cite{73} to measure
the structure functions in the deep-inelastic processes aiming at the fractal
dimensions of the scattered partners. However, it has been done theoretically
on the model level only, and experimental difficulties prevent also the direct
comparison. Therefore the problem of fractal geometry of particles in the
ordinary space-time is still waiting for its solution.

At the very end of the Section, let us mention that no experimental analysis of
the
non-integer rank moments has been performed yet, and their discussion in
Sections
2 and 3 should be interpreted as a Chekhov rifle in the play which is still
pretty far from being finished.

\Section{Brief discussion of other QCD effects}

Theoretical foundation of quantum chromodynamics as the non-abelian gauge
theory describing the interaction of quarks and gluons has been firmly
established though the confinement problem has not been resolved yet.  The
prediction of the asymptotical freedom, i.e. of the weaker interaction at small
distances, allows to apply the well-developed methods of the perturbation
theory.
In parallel, there are lattice calculations, some symmetry properties are
considered, the potential model approach is elaborated. The hadronization of
quarks and gluons is described in some models which are compared to experiment
if the Monte Carlo calculations are done. The whole arsenal of theoretical
methods
provides many predictions and describes a lot of effects observed
experimentally.
Among them there are such successes of quantum chromodynamics as the ratio of
the total cross sections of $e^{+}e{-}$-annihilation to hadrons  and to the
muon
pair, predictions of the sum rules, which interrelate quarks of various
species, detailed description of the properties of heavy quarkonia, some
relations for polarized particles etc.

The achievements of the perturbative methods have been connected, first, with
the hard processes of particle interactions where the transferred momenta are
high. Those are the $e^{+}e^{-}$-annihilation at high energies, deep-imelastic
scattering of leptons and neutrino on hadrons, hadronic processes with large
transverse momenta, heavy quark production, Drell-Yan process (production of
muon pairs with large invariant mass). The universality of the structure
functions as applied in various reactions, the correspondence of the lowest
approximation of quantum chromodynamics to the parton model results including
scaling property and subsequent violation of it in the higher  order terms
due to the running coupling constant were noticeable landmarks of the theory
evolution and its comparison to experiment.

The interference effects were studied somewhat later. We shall mention them at
some length. The notion of coherence plays a predominant role in many outcomes
of theoretical calculations \cite{74,75}. It is a coherence which allows to
apply the
evolution equations for generating functions considered above on the
probabilistic level to description of jets. One of the most remarkable
predictions is the so-called "hump-backed" plateau \cite{76}. The coherence is
especially important for various correlation characteristics (energy flows,
multiplicities etc. \cite{10,35,71}). Recently, the so called string \cite{77}
or drag \cite{78}-effect has been studied experimentally. The theory
predicts that in three-jet $e^{+}e{-}$-events
there is a depletion of particles in the region between two quarks because
they are "dragged" by the gluon jet while there is no such effect if the photon
is emitted in the same direction. It is explained as the destructive
interference
in between the two quarks. The analogous effects have been predicted in
reactions  of creation of photons, muon pairs, heavy bosons with high
transverse  momenta.

Another effect of suppression of hadron multiplicity in reactions with heavy
quarks has been observed in \cite{79}. If compared to the analogous process
with
production  of light quarks, the so called companion multiplicity of
secondaries
in heavy quark production is several charged particles smaller. By that one
means the particles which appear during the hadronization of bremsstrahlung
gluons emitted by the initial quark. The origin of the effect lies
in large masses of heavy quarks. In that case, as is known from scattering
theory \cite{80}, the radiation of vector particles (in our case, gluons)
in the direction  of the heavy (or fast-decaying) parent is suppressed and the
total intensity is reduced. Thus, this observation confirms the vector nature
of gluons once again. It would be interesting to measure the angular
distribution of such events which should reveal
the ring-like structure of the angular distribution of "bremsstrahlung" hadrons
\cite{81,82} with a dead cone inside \cite{83}.

In general, the idea of angular  ordering of quark-gluon jets
due to interference effects is very fruitful and gives rise to equations
for generating functions of multiplicity distributions considered above.

\Section{Conclusions}

There is the principal difference between the effects described in the previous
section and the results about multiplicity distributions compiled in the paper.
The former ones are somehow determined by the hardness of the processes while
the multiplicity distributions are mostly related to soft particles appearing
at
the latest stages of the development  of the cascade. Therefore the success of
the approach substantially extends the range of the pretentious attempts
of quantum chromodynamics. The higher order approximations for the running
coupling constant and the exact solution for the fixed coupling show
qualitatively new features  of multiplicity distributions compared to the
double logarithmic approximation. From the physical
point of view they should take into account the softer stages of the
parton cascade. The qualitative features predicted for partons happen to be
valid for hadrons produced both in $e^{+}e^{-}$ and $hh$ processes. It prompts
the speculation about similarity of the production mechanisms in both cases and
about applicability of higher order perturbative results to description of
soft stages.

Evolution of the attitude to extension of the validity region of the
perturbative approach can be traced in the history of the problem as
described in the introduction. The initial excitement stimulated by
predictions of the energy increase of the average multiplicity and of
the KNO--scaling independent of the coupling constant gave a way to some
depression evoked by the extremely wide multiplicity distribution
predicted theoretically though
it was soon stated that the correction terms were rather large. Now it
becomes completely clear that the next-to-next-to leading corrections are
important also and it is possible to take them into account in a correct way.
The results reveal approximate $F$- (or KNO-) scaling with dependence on the
coupling constant. New peculiarities in the behaviour of the moments of
multiplicity distributions have been predicted and confirmed by experiment.
The evolution of distributions in smaller phase space regions has been
described also. The former discrepancy in the value of the ratio of the
average multiplicities in gluon and quark jets has been practically solved.
The influence of the higher-order corrections on the energy increase of
the average multiplicity has been shown. All these improvements point out
in the direction of better agreement with experiment.

In combination with predictions of the inclusive spectra and various
correlation functions, the above results on the multiplicity distributions
tell us that quantum chromodynamics may be successfully applied to predict
quite special qualitative features of soft processes when considered in
higher orders. Surely, one needs the hadronization scheme and
Monte Carlo calculation to proceed to the quantitative comparison with
experiment (it is often substituted by the assumption about the local parton-
hadron duality). Unfortunately, they suffer from abundance of fitted parameters
which are hard to control sometimes. That is why the analytical predictions
of new qualitative features and effects are of uppermost value. The progress in
that direction, described at some length above, gives some hope for further
success.

\vspace{2mm}
{\bf \{Acknowledgements}}
\vspace{2mm}

I am very much indebted to Yu.L. Dokshitzer, G. Gianini, R.C. Hwa,
B.B. Levtchenko and V.A. Nechitailo with whom I collaborated on the subject.

This work was supported by Russian Fund for Fundamental Studies (grant 93-02-
3815) and by International Science Foundation (grant M5V000).

{\bf Figure Captions}

Fig.1  Fractional factorial moments [15] of Poisson (e) and negative binomial
       distributions (a-d correspond to various values of $k$) with the
       average multiplicity $\langle n \rangle $=2. The amplitude of
oscillations
       decreases strongly at larger $\langle n \rangle $ and smaller $k$.

Fig.2  Moments of the negative binomial distribution [21] for $k$=5 and 10
       calculated for integer values of $q$. The curves are drawn to guide
       the eye.\\
       (a) $\ln F_{q}$ , (b) $\ln K_{q}$ , (c) $\ln H_{q}$ .

Fig.3  Moments of fixed multiplicity distribution [21] for $n_{0}$=10
calculated
       for integer values of $q$. The curves are drawn to guide the eye. \\
       (a) $F_q$ , (b) $K_q$ , $\ln \vert K_q\vert $ , (c) $H_q$ , $\ln \vert
       H_{q}\vert $ .

Fig.4  (a) The ratio of factorial moments as derived from eq.(\ref{58}) to the
       asymptotical values of eq.(\ref{59}) (the inset) and the similar
       ratio as derived from the KNO--curve (see Fig.5) and eq.(\ref{59})
       (the main part).  \\
       (b) The ratio $H_q$ obtained from the KNO--curve (see Fig.5) (the solid
       line) as compared to its NBD-counterpart for $k$=7 (the dashed line).\\
       I am indebted to B.B. Levtchenko who provided this Figure specially for
       the review.

Fig.5  The modified KNO--function [12] (solid curve) for $\gamma $=0.4 is much
narrower
       than the lowest order distribution (the dashed line). The negative
binomial
       distribution with $k$=7 is also shown for comparison (dots).

Fig.6  The ratio $H_q$ as a function of $q$ [39] reveals "quasi-oscillations"
in
       higher order perturbative QCD (the curve is drawn for energies of
$Z^{0}$).
       The first minimum is slightly shifted (compared to gluodynamics) to
$q$=4.

Fig.7  $Q$-dependence of the anomalous dimension $\gamma $ [41] (solid lines)
in
       case of the running coupling $\gamma _{0}$ (dashed lines) for different
       number (shown near the lines) of active flavours.

Fig.8  $y$-dependence of the average multiplicity for running (solid lines) and
       fixed (dashed lines) coupling [41] with the number of active flavours
       $n_f$=3,4,5. The arrow marks the $y_{Z^0}$ location.

Fig.9  $\gamma $ vs $\gamma _{0}$ for $n_f$=3,4,5 [42].

Fig.10 $r$ vs $\gamma _{0}$ for $n_f$=3,4,5 [42].

Fig.11 $\gamma $ vs $\ln Q$ for $n_f$=3,4,5 and $Q=M_{Z}/m  [42]\;\;
(m=1,2,4,8)$.

Fig.12 Moments of multiplicity distribution in fixed-coupling QCD for
       $\gamma _{0}$=0.48, $n_f$=5 [21]. \;\; (a) $\ln F_{q} , \ln \Phi _{q}$ ,
       (b) $\ln \vert K_{q}\vert , \ln \vert \Psi _{q}\vert $ .

Fig.13 Ratio $H_q$ of gluon-jet distribution in fixed-coupling QCD for
       $\gamma _{0}$=0.48 , $n_f$=5 [21]. \;\; (a) $H_q$ , (b) $\ln \vert
H_{q}\vert $ .

Fig.14 Ratio $\eta _{q}$ of quark-jet distribution in fixed-coupling QCD for
       $\gamma _{0}$=0.48 , $n_f$=5 [21]. \;\; (a) $\eta _{q}$ , (b) $\ln \vert
       \eta _{q} \vert $ .

Fig.15 Ratio $r$ of average multiplicities in gluon and quark jets [48].\\
       Analytic results are shown by upper curves. The results obtained from
       the Herwig Monte Carlo at the parton and hadron levels are also shown.

Fig.16 $H_q$ vs $q$ in $e^{+}e^{-}$ data of DELPHI collaboration at 91 GeV
[51].

Fig.17 $H_q$ vs $q$ in $p\bar {p}$ data of UA5 collaboration at 546 GeV [51].

Fig.18 $H_q$ vs $q$ in $e^{+}e^{-}$ data in a wide energy interval [51]
       (experimental groups are ordered as in Table 2, i.e. the energy
       increases from top to bottom). On the left the lowest orders are shown
       in the log scale, on the right the higher orders in the linear scale.

Fig.19 $H_q$ vs $q$ in $pp$(and $\bar p$ data in a wide energy interval [51]
       (the comments are the same as in Fig.18).

Fig.20 Emission of the gluon (wavy lines) jet by the quark (the solid line)
[38].


\begin{thebibliography} {85}
\bibitem{1}
I.V. Andreev, Chromodynamics and hard processes at high energies, Moscow,
Nauka, 1981 (in Russian).
\bibitem{2}
B.L. Ioffe, L.N. Lipatov and V.A. Khoze, Deep-inelastic processes, Moscow,
Energoatomizdat, 1983 (in Russian).
\bibitem{3}
F.J. Yndurain, Quantum chromodynamics, N.-Y.-Berlin-Heidelberg-Tokyo,
Springer Verlag, 1983.
\bibitem{4}
M.B. Voloshin and K.A. Ter-Martirosyan, Theory of gauge interactions of
elementary particles, Moscow, Energoatomizdat, 1984 (in Russian).
\bibitem{5}
Yu.L. Dokshitzer, V.A. Khoze, A.H. Mueller and S.I. Troyan, Basics of
perturbative QCD, Gif-sur-Yvette, Editions Frontieres, 1991.
\bibitem{6}
A Giovannini and L Van Hove, Z. Phys. C30 (1986) 391; Acta Phys. Pol. B19
(1988) 495;917;931.
\bibitem{7}
Z. Koba, H.B. Nielsen and P. Olesen, Nucl. Phys. B40 (1972) 317.
\bibitem{8}
Ya.I. Azimov, Yu.L. Dokshitzer, V.A. Khoze and S.I. Troyan, Z. Phys. C27 (1985)
65.
\bibitem{9}
D. Amati and G. Veneziano, Phys. Lett. B83 (1979) 87.
\bibitem{10}
A. Bassetto, M. Ciafaloni and G. Marchesini, Phys. Rep. C100 (1983) 201.
\bibitem{11}
A. Bassetto, M. Ciafaloni and G. Marchesini, Nucl. Phys. B163 (1980) 477.
\bibitem{12}
Yu.L. Dokshitzer, Phys. Lett. B305 (1993) 295.
\bibitem{13}
I.M. Dremin, Phys. Lett. B313 (1993) 209.
\bibitem{14}
E.A. DeWolf, I.M. Dremin and W. Kittel, Phys. Rep. (1994).
\bibitem{15}
E.M. Friedlander and I. Stern, Preprint LBL-31354, 1991 (unpublished).
\bibitem{16}
K. Oldham, The fractional calculus, Orlando, Academic Press, 1974, p.60; \\
B. Ross, Fractional calculus and its applications, in Lecture Notes in
Mathematics, Berlin, Springer Verlag, v.457 (1975) 1.
\bibitem{17}
I.M. Dremin, JETP Lett. 60 (1994) (to be published).
\bibitem{18}
F. Cuypers and K. Tesima, Z. Phys. C54 (1992) 87.
\bibitem{19}
Yu.L. Dokshitzer, V.A. Khoze and S.I. Troyan, in {\it Perturbative QCD},
ed. A.H. Mueller, Singapore, World Scientific, 1989.
\bibitem{20}
I.M. Dremin, Mod. Phys. Lett. A8 (1993) 2747.
\bibitem{21}
I.M. Dremin and R.C. Hwa, Phys. Rev. D (1 June 1994).
\bibitem{22}
B. Andersson, G. Gustafson and T. Sjostrand, Phys. Lett. B94 (1980) 211.
\bibitem {23}
B. Andersson, G. Gustafson, G. Ingelman and T. Sjostrand, Phys. Rep. 97 (1983)
33.
\bibitem{24}
A. Capella, U. Sukhatme, C.I. Tan and J. Tran Thanh Van, Phys. Lett. B81
(1979) 68.
\bibitem{25}
A. Capella and J. Tran Thanh Van, Phys. Lett. B93 (1980) 146;
Z. Phys. C10 (1981) 249.
\bibitem{26}
A.B. Kaidalov, Phys. Lett. B116 (1982) 459.
\bibitem{27}
A.B. Kaidalov and K.A. Ter-Martirosyan, Phys. Lett. B117 (1982) 247;
Sov. J. Nucl. Phys. 39 (1984) 979; 40 (1984) 135.
\bibitem{28}
I.M. Dremin and A.M. Dunaevskii, Phys. Rep. 18 (1975) 159.
\bibitem{29}
R. Ugoccioni, A. Giovannini and S. Lupia, in Proc. 23 Int. Symp. on
Multiparticle Dynamics, Aspen, USA, 1993 (to be published).
\bibitem{30}
B.B. Levtchenko and  A.V. Shumilin, Zs. Phys. (1994) (to be published).
\bibitem{31}
F. Bianchi, A. Giovannini, S. Lupia and R. Ugoccioni, Z. Phys. C58 (1993) 71.
\bibitem{32}
Yu.L. Dokshitzer and M. Olsson, Nucl. Phys. B396 (1993) 137.
\bibitem{33}
E.D. Malaza and B.R. Webber, Nucl. Phys. B267 (1986) 702.
\bibitem{34}
J.B. Gaffney and A.H. Mueller, Nucl. Phys. B250 (1985) 109.
\bibitem{35}
F. Cuypers and K. Tesima, Z. Phys. C52 (1991) 69; C54 (1992) 87.
\bibitem{36}
S. Catani, Yu.L. Dokshitzer, F. Fiorani and B.R. Webber, Nucl. Phys. B377
(1992)
445; B383 (1992) 419.
\bibitem{37}
M. Olsson and G. Gustafson, Nucl. Phys. B406 (1993) 293.
\bibitem{38}
Yu.L. Dokshitzer and I.M. Dremin, Nucl. Phys. B402 (1993) 139.
\bibitem{39}
I.M. Dremin and V.A. Nechitailo, JETP Lett. 58 (1993) 945.
\bibitem{40}
I.M. Dremin, B.B. Levtchenko and V.A. Nechitailo, Sov. J. Nucl. Phys. 59
(1994) N6.
\bibitem{41}
I.M. Dremin and V.A. Nechitailo, Mod. Phys. Lett. A9 (1994).
\bibitem{42}
I.M. Dremin and R.C. Hwa, Phys. Lett. B324 (1994) 477.
\bibitem{43}
A.H. Mueller, Nucl. Phys. B241 (1984) 141.
\bibitem{44}
E.D. Malaza and B.R. Webber, Phys. Lett. B149 (1984) 501.
\bibitem{45}
W.J. Marciano, Phys. Rev. D29 (1984) 580.
\bibitem{46}
Yu.L. Dokshitzer, V.A. Khoze, A.H. Mueller and S.I. Troyan,
Rev. Mod. Phys. 60 (1988) 373.
\bibitem{47}
G. Marchesini, B.R. Webber et al., Comp. Phys. Comm. 67 (1992) 465.
\bibitem{48}
J.W. Gary, Preprint UCRHEP-T116 (1993).
\bibitem{49}
OPAL coll., P.D. Acton et al., Z. Phys. C58 (1993) 387.
\bibitem{50}
CLEO coll., M.S. Alam et al., Phys. Rev. D46 (1992) 4822.
\bibitem{51}
G. Gianini et al., In Proc. 23 Int. Symp. on Multiparticle Dynamics, Aspen,
USA, 1993; Singapore, World Scientific (to be published).
\bibitem{52}
TASSO coll., W. Braunschweig et al., Z. Phys. C45 (1989) 193.
\bibitem{53}
HRS coll., M. Derrick et al., Phys. Rev. D34 (1986) 3304.
\bibitem{54}
ALEPH coll., D. Decamp et al., Phys. Lett. B273 (1991) 181.
\bibitem{55}
DELPHI coll., P. Abreu et al., Z. Phys. C50 (1991) 185.
\bibitem{56}
L3 coll., B. Adeva et al., Z. Phys. C55 (1992) 39.
\bibitem{57}
OPAL coll., P.D. Acton et al., Z. Phys. C53 (1992) 539.
\bibitem{58}
V. Ammosov et al., Phys. Lett. B42 (1972) 519.
\bibitem{59}
W.M. Morse et al., Phys. Rev. D15 (1977) 66.
\bibitem{60}
G. Charlton et al.,  Phys. Rev. Lett. 29 (1972)515.
\bibitem{61}
A. Firestone et al., Phys. Rev. D9 (1974) 2080.
\bibitem{62}
E743 coll., R. Ammar et al., Phys. Lett. B178 (1986) 124.
\bibitem{63}
A. Breakstone et al., Phys. Rev. D30 (1984) 30.
\bibitem{64}
UA5 coll., R.E. Ansorge et al., Z. Phys. C43 (1989) 357.
\bibitem{65}
UA5 coll., G.J. Alner et al., Phys. Rep. 154 (1987) 247.
\bibitem{66}
A. Bialas and R. Peschanski, Nucl. Phys. B273 (1986) 703.
\bibitem{67}
G. Gustafson and A. Nilsson, Z. Phys. C52 (1991) 533; Nucl. Phys. B355 (1991)
106.
\bibitem{68}
W. Ochs and J. Wosiek, Phys. Lett. B289 (1992) 159.
\bibitem{69}
W. Ochs and J. Wosiek, Phys. Lett. B (1993); Preprint MPI-PH-93-30.
\bibitem{70}
Ph. Brax, J.L. Meunier and R. Peschanski, Nucl. Phys. B (1993); Preprint
INLN-93-01.
\bibitem{71}
Yu.L. Dokshitzer, G. Marchesini and G. Oriani, Nucl. Phys. B387 (1992) 675.
\bibitem{72}
M.I. Polikarpov, Phys. Lett. B236 (1990) 61.
\bibitem{73}
I.M. Dremin and B.B. Levtchenko, Phys. Lett. B292 (1992) 155.
\bibitem{74}
A.H. Mueller, Phys. Lett. B104 (1981) 161.
\bibitem{75}
G. Marchesini and B.R. Webber, Nucl. Phys. B238 (1984) 1.
\bibitem{76}
Ya.I. Azimov, Yu.L. Dokshitzer, V.A. Khoze and S.I. Troyan, Z. Phys.
C31 (1986) 213.
\bibitem{77}
B. Andersson, G. Gustafson and T. Sjostrand, Phys. Lett. B94 (1980) 211.
\bibitem{78}
Ya.I. Azimov, Yu.L. Dokshitzer, V.A. Khoze and S.I. Troyan, Phys. Lett.
B165 (1985) 147.
\bibitem{79}
B.A. Schumm, Yu.L. Dokshitzer, V.A. Khoze and D.S. Koetke, Phys. Rev. Lett.
69 (1992) 3025.
\bibitem{80}
I.M. Dremin, JETP Lett. 34 (1981) 617.
\bibitem{81}
I.M. Dremin, JETP Lett. 30 (1979) 152.
\bibitem{82}
A.V. Apanasenko, N.A. Dobrotin, I.M. Dremin and K.A. Kotelnikov,
JETP Lett. 30 (1979) 157.
\bibitem{83}
Yu.L. Dokshitzer, V.A. Khoze and S.I. Troyan, J. Phys. G17 (1991) 1481, 1602.
\end{thebibliography}
\end{document}